\documentclass[journal]{IEEEtran}
\usepackage{xcolor}
\usepackage{soul}
\soulregister\cite7
\soulregister\ref7 
\definecolor{Skyblue}{RGB}{135,206,235} 
\definecolor{Black}{RGB}{0, 0, 0}

\usepackage{cite}

\ifCLASSINFOpdf
  \usepackage[pdftex]{graphicx}
   \graphicspath{{../pdf/}{../jpeg/}}
   \DeclareGraphicsExtensions{.pdf,.jpeg,.png}
\else
   
   \usepackage[dvips]{graphicx}
   \graphicspath{{../eps/}}
\fi
\usepackage{amsmath}
\usepackage{amsfonts} 
\usepackage{algorithmic}
\usepackage{multirow}

\newtheorem{remark}{Remark}

\usepackage{array}
\usepackage{fixltx2e}

\usepackage{stfloats}
\usepackage{diagbox}
\usepackage{url}

\usepackage{tabularx,booktabs}
\usepackage{makecell}
\usepackage{caption}
\usepackage{subcaption}
\begin{document}
%
\title{Performance Analysis of Uplink Rate-Splitting Multiple Access with Hybrid ARQ }
%
%
%
\author{{Yuanwen Liu,   
       Bruno~Clerckx,~\IEEEmembership{Fellow,~IEEE,}
      Petar~Popovski,~\IEEEmembership{Fellow,~IEEE,}}
 \thanks{Yuanwen Liu is with the Department of Electrical and Electronic Engineering, Imperial College London, London SW7 2AZ, U.K. (e-mail: y.liu21@imperial.ac.uk).}
 \thanks{Bruno Clerckx is with the Department of Electrical and Electronic Engineering, Imperial College London, London SW7 2AZ, U.K. (e-mail:b.clerckx@imperial.ac.uk).}
\thanks{Petar Popovski is with the Department of Electronic Systems, Aalborg University, 9220 Aalborg, Denmark (e-mail: petarp@es.aau.dk).}}
%
%

\markboth{Journal of \LaTeX\ Class Files,~Vol.~14, No.~8, August~2021}%
{Shell \MakeLowercase{\textit{et al.}}: A Sample Article Using IEEEtran.cls for IEEE Journals}

%

\IEEEpubid{0000--0000/00\$00.00~\copyright~2021 IEEE}

\maketitle
\begin{abstract}
Rate-splitting multiple access (RSMA) has attracted a lot of attention as a general and powerful multiple access scheme. In the uplink, instead of encoding the whole message into one stream, a user can split its message into two parts and encode them into two streams before transmitting a superposition of these two streams. The base station (BS) uses successive interference cancellation (SIC) to decode the streams and reconstruct the original messages. Focusing on the packet transmission reliability, we investigate the features of RSMA in the context of hybrid automatic repeat request (HARQ), a well-established mechanism for enhancing reliability. This work proposes a HARQ scheme for uplink RSMA with different retransmission times for a two-user scenario and introduces a power allocation strategy for the two split streams. The results show that compared with non-orthogonal multiple access (NOMA) and frequency division multiple access (FDMA), RSMA outperforms them in terms of error probability and power consumption. The results show that RSMA with HARQ has the potential to improve the reliability and efficiency of wireless communication systems.
\end{abstract}


%
\IEEEpeerreviewmaketitle

\section{Introduction}
%
%
%
%


\IEEEPARstart{6}{G} has drawn the attention of academia and industry due to its ability to offer new services requiring higher throughput, ultra reliability, low latency, and massive connectivity for everything. In current communication networks, non-contention access methods, such as orthogonal time–frequency division multiple access, are popular in communication systems. However, these methods are not suitable for applications involving massive low-power devices connected to a common BS, because they cannot accommodate numerous devices orthogonally with limited resources. This calls for rethinking multiple access (MA) techniques \cite{9390169}.

RSMA is a flexible, universal, efficient, robust, and reliable MA technique that generalizes and outperforms numerous seemingly unrelated MA schemes, including orthogonal multiple access (OMA), NOMA, multicasting, and space division multiple access (SDMA). It has been shown that RSMA finds numerous new applications in 6G \cite{9831440,https://doi.org/10.48550/arxiv.2209.00491}. In the downlink, each message is split into two parts, a common part and a private part, and all the common parts are encoded into one common stream while the private parts are independently encoded into private streams. The BS precodes all common and private streams and transmits these superposed streams to the users. At the receiver, each user uses SIC to decode the common stream first and then decodes its own private stream. The message split enables RSMA to partially decode interference and partially treat the remaining interference as noise by adjusting the power allocation between common streams and private streams properly. Consequently, RSMA softly bridges SDMA and NOMA as a more flexible and general technique and outperforms them \cite{article,8907421}. RSMA has not only higher spectrum efficiency, but also higher energy efficiency for various applications \cite{8491100,7738598,9145189,8846706}. It can also support low latency services \cite{9217326,9562192,downlinkFBL,Fairness_FBL}, and it has the potential to interplay with other techniques, such as reconfigurable intelligent surfaces \cite{rsma-ris}, joint sensing and communications \cite{9832622}, integrated terrestrial and non-terrestrial networks \cite{https://doi.org/10.48550/arxiv.2209.00491}, etc. RSMA is also a powerful and robust strategy for multi-user multiple-input multiple-output \cite{9663192}. These advantages show that RSMA can be a competitive multiple access strategy for future networks \cite{9348672,9451194,9831440}. 

Uplink RSMA is studied in \cite{485709,915637,9257190,9064705,CR-RSMA,8171078,RSMA_semi_grant_free,uplinkFBL,9643016,networkslicing_RSMA} and was first proposed in \cite{485709}. The users split their messages into two parts and encode them into two streams. This can be seen as adding a virtual user, and users transmit a superposition of these two streams. BS uses SIC to decode the streams and then reconstructs the messages. The increasing number of streams provides a more flexible decoding order and enables RSMA to achieve all the boundary points of the capacity region without time-sharing \cite{9257190}. RSMA can also improve user fairness, outage performance \cite{9064705,CR-RSMA}, simplify the implementation by avoiding complex user pairing \cite{8171078}, increase the connectivity in semi-grant-free transmission \cite{RSMA_semi_grant_free}, and also perform well in finite blocklength regime \cite{uplinkFBL}. Although NOMA with time sharing can achieve the same capacity region as RSMA, it needs multiple time slots \cite{9831440} and induces communication overhead and latency, which can be more complex than RSMA \cite{9257190}. Therefore, RSMA is a promising technique for any service which needs grant-free and is required intermittently, i.e., ultra-reliable low-latency communication (URLLC) and massive machine-type communication (mMTC) \cite{8476595}. \cite{9643016,networkslicing_RSMA} showed that RSMA can outperform orthogonal multiple access (OMA) and NOMA in some heterogeneous services coexistence scenarios. Consequently, uplink RSMA shows the potential to be applied in future communication systems.
\IEEEpubidadjcol
Given the numerous promising properties, it is worth exploring integrating RSMA with other techniques to improve reliability. One critical mechanism to enhance reliability is HARQ. It also increases diversity and improves the efficiency of packet-based transmission \cite{9831440}. Thus, it is a crucial technique to meet the expectations in future networks \cite{9152144,9254087}. This mechanism is based on forward error correction (FEC) and automatic repeat request (ARQ) \cite{930931}. Generally, there are three types of HARQ: Type I HARQ, HARQ with chase combining (CC), and HARQ with incremental redundancy (IR). Type I HARQ uses the packet in the current round to decode, while HARQ with CC and HARQ with IR can decode packets in different rounds jointly with maximal ratio combining (MRC). In HARQ with CC, the packets are identical in each retransmission turn, and the receiver uses MRC to combine messages in every turn; while in HARQ with IR, additional parity bits are transmitted in each retransmission turn \cite{6487360}. HARQ has been extensively studied with both OMA \cite{6487360,6502167} and NOMA \cite{9254087,8884744,7501524,9089002,8408492,e23070880,noma-harq}. For OMA, \cite{6487360} optimized the transmission power for a given outage probability with limited retransmission times for the CC scheme, and \cite{6502167} maximized the throughput in the IR scheme. For NOMA, \cite{8884744,7501524,9089002,8408492} focused on the downlink scenario and showed that NOMA with HARQ can improve the outage probability and energy efficiency. \cite{9254087,e23070880,noma-harq} studied NOMA with HARQ in uplink, and showed that it can reduce the latency and save power consumption. Specifically, \cite{9254087,e23070880,noma-harq} showed NOMA with HARQ has promising performances for short packets, which means it is suitable for services requiring low latency. Performing HARQ orthogonally would not be able to accommodate numerous users with limited resources. It also increases the latency, because unaccommodated users need to wait in the queue. Clearly, allocating resources and performing HARQ non-orthogonally can improve the efficiency of systems and reduce the waiting time. Besides, BS can keep the previous unsuccessful copies of messages, so using a full payload to retransmit messages again would be wasteful. Meanwhile, fixed-size resources are preferred, so it is not practical to adjust the size of retransmission \cite{9254087}. However, HARQ with NOMA is doomed to face the problem that users sharing the same resources interfere with each other. When users require the service at a high rate, even the user with a better condition may not be decoded due to interference from the other users, and it is almost impossible to decode the user with worse channels. Then in the retransmission turn, if the channel is still not good enough, an additional retransmission turn would be needed. Therefore, this could increase the latency \cite{8972353}.

\begin{figure}
     \centering
     \begin{subfigure}[]{0.5\textwidth}
         
         \includegraphics[width=8cm]{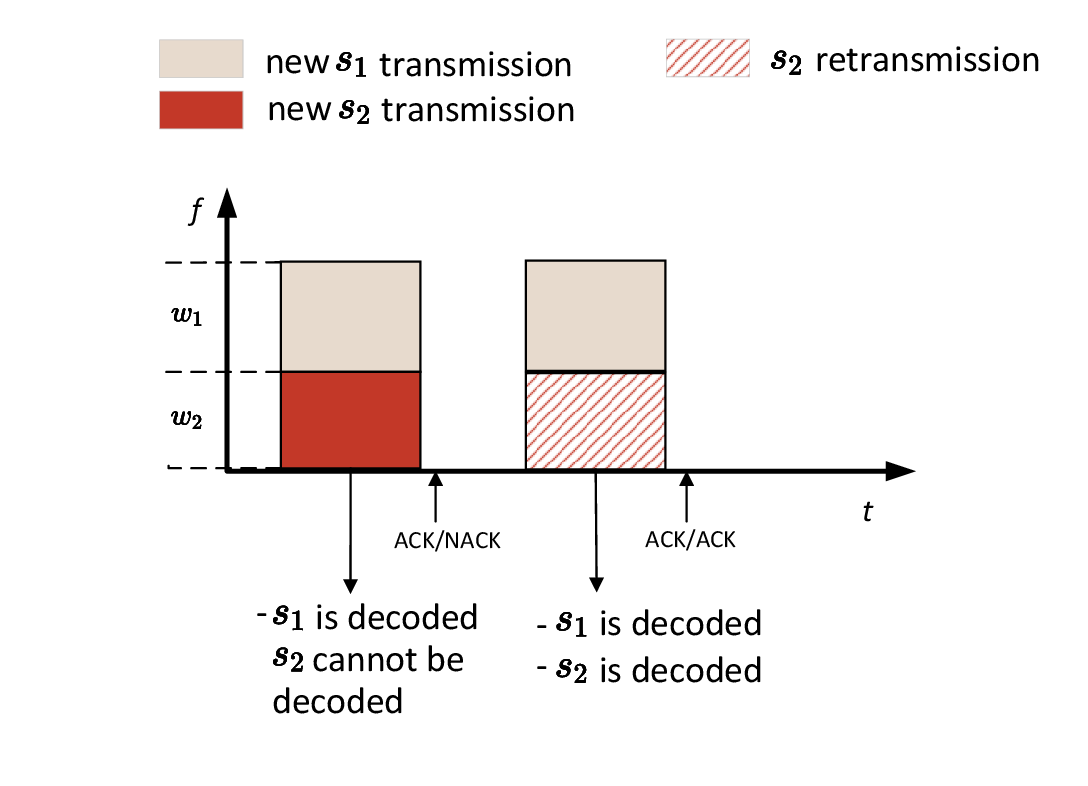}
         \caption{FDMA}
         \label{fig:FDMA}
     \end{subfigure}

     \quad
     
     \begin{subfigure}[]{0.5\textwidth}
         \includegraphics[width=8cm]{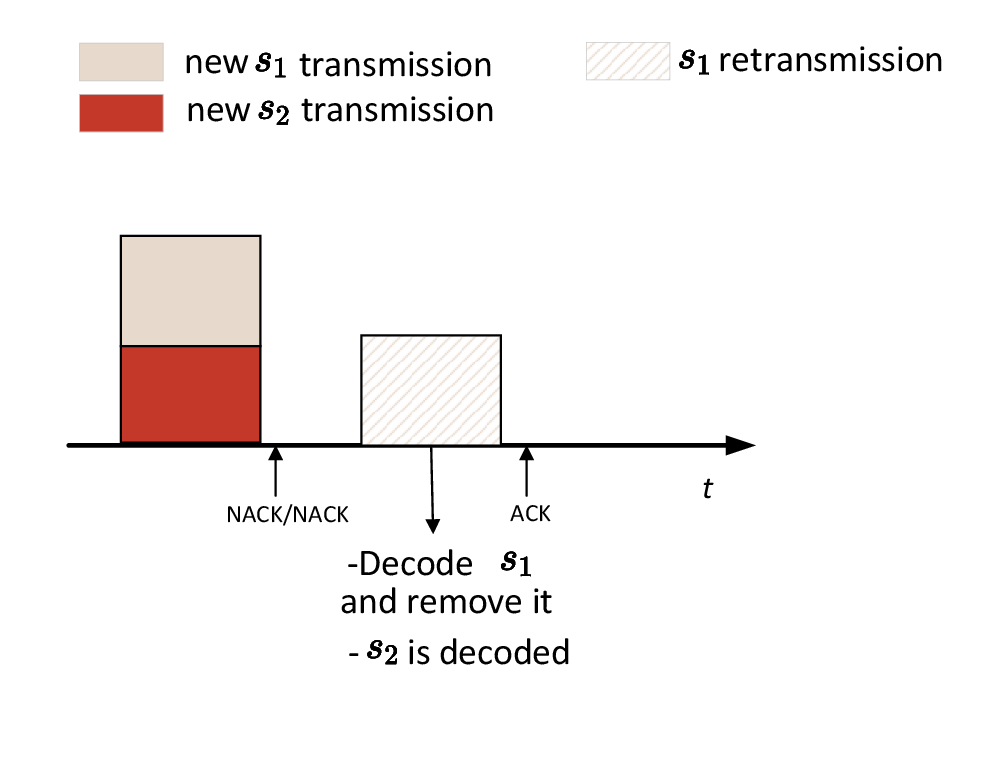}
         \caption{NOMA}
         \label{fig:NOMA}
     \end{subfigure}

    \quad
     
     \begin{subfigure}[]{0.5\textwidth}
        
         \includegraphics[width=8cm]{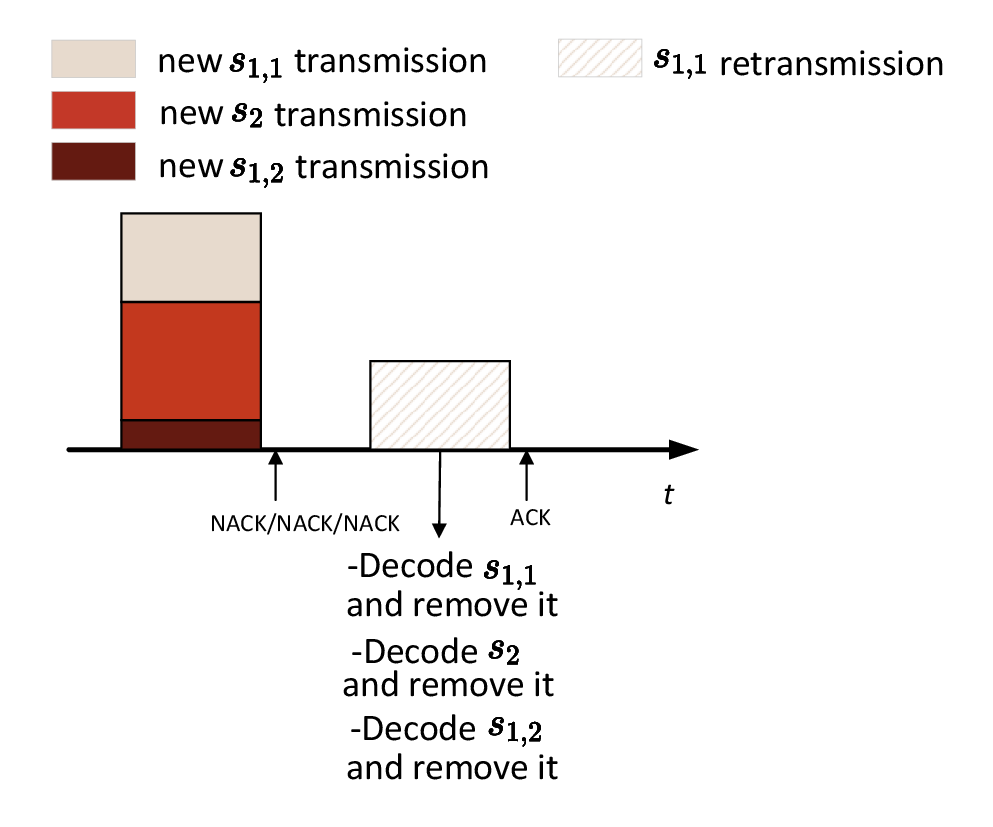}
         \caption{RSMA}
         \label{fig:RSMA}
     \end{subfigure}
        \caption{The retransmission schemes of FDMA, NOMA and RSMA}
        \label{fig:a toy example}
\end{figure}

RSMA can manage interference better by bringing more streams, and it provides higher flexibility for transmission. \cite{loli2023hybrid} studied HARQ scheme for downlink RSMA, and showed it increased the success probability of transmission, while HARQ for uplink RSMA has not been investigated yet. Since RSMA splits its message into two parts, the number of transmitted streams increases, so the number of retransmitted streams may also increase. This increase can bring more complexity during the retransmission process. This motivates us to investigate how we can design a HARQ scheme tailored for RSMA to mitigate the additional complexity. This work focuses on HARQ for uplink RSMA. Fig. \ref{fig:a toy example} shows a toy example of RSMA retransmission scheme and compares it with FDMA and NOMA. Let $s_1$ and $s_2$ denote the signals transmitted by user 1 and user 2, respectively. $t$ denotes time and $f$ denotes frequency. Fig. \ref{fig:FDMA} shows FDMA, and user 1 and user 2 occupy non-overlapping frequency bands $w_1$ and $w_2$, respectively. If one message cannot be decoded, this message will be retransmitted and will not interfere with the other user. In Fig. \ref{fig:FDMA}, at first $s_2$ cannot be decoded and user 2 receives a negative acknowledgement (NACK). Then the retransmitted $s_2$ and new $s_1$ are sent in the next time slot, and both users receive acknowledgement (ACK). In Fig. \ref{fig:NOMA}, a NOMA retransmission scheme is shown. Two users can share the same time-frequency resource, and BS uses SIC to decode the two messages. We assume that user 1 has a better channel condition and $s_1$ will be decoded first. If $s_1$ fails to be decoded, neither $s_2$ can be decoded. But we do not need to always retransmit all the failed messages, because sometimes $s_2$ can be decoded directly once $s_1$ is decoded. Therefore, sometimes only $s_1$ is retransmitted, and after BS decodes $s_1$ successfully, $s_2$ is decoded. Fig. \ref{fig:RSMA} shows the retransmission scheme of RSMA. $s_{1,1}$ and $s_{1,2}$ denote the two split streams of user 1. BS needs to decode both $s_{1,1}$ and $s_{1,2}$ to obtain the message of user 1. We assume that user 1 has a better channel condition and the decoding order is $s_{1,1}$, $s_2$, and $s_{1,2}$. Similarly, if $s_{1,1}$ fails to be decoded, neither $s_2$ and $s_{1,2}$ can be decoded. But sometimes only $s_{1,1}$ needs to be retransmitted, and $s_2$ and $s_{1,2}$ will be decoded once BS decodes $s_{1,1}$ successfully. Thus, compared to FDMA, RSMA has a higher spectrum efficiency; while compared to NOMA, RSMA can retransmit one of the split streams and has the potential to consume less energy.

We note that we make two assumptions here. The two users require the same service, so they have the same reliability requirements. The strong user splits its message into two parts and encodes them into two streams, and the weak user encodes the whole message and transmits only one stream. The contributions of this work are summarized below:

\begin{itemize}
    \item We propose a retransmission scheme for uplink RSMA. For a usual HARQ scheme, the failed steams will be retransmitted. Since RSMA splits its messages into two parts, the number of transmitted streams increases. Thus, more streams may need retransmissions if they fail to be decoded. This increase will bring more overhead and complexity during the retransmission turns. Therefore, an appropriate design  HARQ scheme for RSMA is necessary. In this work, we focus on how to design a suitable HARQ scheme tailored for uplink RSMA. This scheme does not need to retransmit all the failed streams, so it will not bring additional complexity. Only one stream of the strong user and the stream of the weak user are involved in retransmission. The two streams of the strong user do not need to be decoded successively, so BS can decode these two streams and the stream of the weak user alternatively. This decoding order is optimal and it can fully exploit decoding flexibility. For the strong user, the sum rate of the two streams is the achievable rate. Thus, one of the streams can always be decoded successfully with a low rate, and the other stream needs to be retransmitted if the sum rate does not satisfy the desired rate. Therefore, only two of three streams involve retransmissions. To our best knowledge, HARQ with RSMA in uplink has not been studied before.
    
    \item  The power allocation strategy between the two streams of the strong user is introduced in this work. When SIC is used, we assume that if one stream cannot be decoded, the following streams will fail. However, it is not always necessary to retransmit a sequence of failed streams. Sometimes it is possible to only retransmit one of the failed streams. Once this stream is decoded and cancelled after retransmission, BS can continue the SIC process to decode other failed streams. The power allocation between the streams of the strong users can actually decide which stream(s) needs to be retransmitted. This work analyzes how power allocation determines which stream(s) will be retransmitted. Naturally, different power allocation strategies bring different error probabilities of the users, and the error probabilities are given analytically in this paper. Then, the power allocation that brings the lowest error probabilities of the two users is chosen. The analysis of power allocation reveals the essence of the design of RSMA with HARQ. It explains intuitively why this HARQ scheme for RSMA has advantages with insights.
    
    \item The error probabilities and average transmission power per packet for a given rate of RSMA with HARQ are simulated by Monte Carlo and a detailed analysis is presented. The performances are simulated with both CC and IR and different retransmission times are allowed. The results show that although RSMA cannot always decrease the error probability of the strong user, it can decrease the error probability of the weak user dramatically. Since the two users require the same service, it is more critical to enable the weak user to meet the service requirements. In this way, RSMA with HARQ can support a higher rate with the same reliability requirement and retransmission times. It can achieve promising performances regardless of the latency requirement of the service. The results also show that RSMA consumes the least power compared to NOMA and FDMA, because it can mitigate the effect of interference between users. Besides, the strong user can retransmit one stream instead of two to save energy. Hence, RSMA has the potential to be applied to services requiring low power consumption.
\end{itemize}

The organization of the rest of the paper is summarized below. Section \ref{sec:2} introduces the system model of RSMA. Section \ref{sec:3} presents HARQ design for RSMA. We use FDMA and NOMA as baselines, and Section \ref{sec_2_4} introduces HARQ for FDMA and NOMA. Section \ref{sec:4} demonstrates the numerical results, and Section \ref{sec:5} is the conclusion.

\textit{Notations}: $\mathbb{C}$ denotes the complex numbers set. $\mathcal{CN}(\delta,\sigma^2)$ represents a complex Gaussian distribution with mean $\delta$ and variance $\sigma^2$.

\section{System Model}\label{sec:2}
In this section, we will give a brief introduction to uplink RSMA. We consider the scenario that two single-antenna users transmit to a common BS with a single antenna. The two users share one time-frequency block to operate the same application with RSMA. $L$ time retransmissions are allowed, which means a user can transmit its message up to $L+1$ rounds, so if the message can not be successfully decoded after $L+1$ rounds, it will be dropped. The unsuccessful copies will be buffered at BS.

In uplink RSMA, instead of transmitting the whole message, a user can split its message into two parts and encode them into two independent streams. Users send the superposition of the two streams to BS, and BS uses SIC to decode these streams \cite{485709}. This procedure relies on channel state information (CSI), so we make some assumptions about channel knowledge. We assume that users do not have CSI while BS has perfect CSI. When the users send the request for connection, BS can obtain the CSI of the users and decide who will split the message and the power allocation, and then send back this information to the users during this connection setup process. Users do not adjust transmission power because they do not have CSI, and their transmission power is normalized to 1. Let $h_k \in \mathbb{C}$ denote the channel between user $k$ and BS, and the channel is considered as Rayleigh fading channel and fades independently, $h_{k} \sim \mathcal{CN}(0,\Gamma_k)$, where $\Gamma_k$ is the average channel gain of user $k$. For user $k$, the transmitted signal can be represented as
\begin{equation}
    s_k=\sum_{i=1}^2\sqrt{P_{k,i}}s_{k,i},
\end{equation}
where $s_{k,i}$ is the split stream of user $k$ and $P_{k,i}$ is the power allocated to the stream $s_{k,i}$, which should satisfy the power constraint. The received signal at BS is 
\begin{equation}
    y=\sum_{k=1}^2 h_k s_k +n,
\end{equation}
where $n \sim \mathcal{CN}(0,\sigma_n ^2)$ is the additive Gaussian noise. Without loss of generality, the noise power is normalized to $1$.

Actually, for the two-user case of RSMA, boundary points of the capacity region can be reached by one user splitting the message \cite{9257190}, and which user splits its message and how power is allocated can be decided by BS. We assume user 1 has a better channel condition, and it splits its message into $s_{1,1}$ and $s_{1,2}$, and $s_{1,1}$ is allocated a fraction $\alpha$ of the transmit power, where $\alpha \in [0,1]$. Thus, the received signal at BS is 
\begin{equation}
    y=\sqrt{\alpha} h_1 s_{1,1} + \sqrt{1-\alpha} h_1 s_{1,2}+h_2 s_2 +n.
\end{equation}
We assume that $s_{1,1}$ is the stream decoded first without loss of generality, and the decoding order is $s_{1,1}$, $s_2$ and $s_{1,2}$, which is the optimal order to achieve all the boundary points of the capacity region \cite{9257190}. If one stream cannot be decoded, the streams after it are unlikely decoded. Therefore, we assume that if one stream fails, the decoding process terminates, i.e., if $s_{1,1}$ cannot be decoded, $s_2$ and $s_{1,2}$ will not be decoded. Thus, to obtain the whole message from user 1, BS needs to decode $s_{1,1}$, $s_2$ and $s_{1,2}$; while for user 2 $s_{1,1}$ and $s_2$ need to be decoded. 

\section{HARQ design for RSMA}\label{sec:3}
This section introduces HARQ design for RSMA. Subsection \ref{sec_2_1} introduces how $\alpha$ affects which stream(s) will be retransmitted. $\alpha$ also determines the error probabilities of the two users, and the $\alpha$ which brings the lowest sum of error probabilities is chosen. Subsection \ref{sec_2_2} and subsection \ref{sec_2_3} present how to obtain the error probabilities with CC and IR, respectively.  Subsection \ref{sec_2_4} presents the HARQ schemes for FDMA and NOMA.

\subsection{How $\alpha$ Determines Which Stream Needs Retransmissions}\label{sec_2_1}
We assume that at the BS if one stream fails, the decoding process terminates, i.e. if $s_{1,1}$ cannot be decoded, $s_2$ and $s_{1,2}$ will not be decoded. In the usual HARQ scheme, failed streams will be transmitted again till they are decoded or consume all the retransmission turns. Since the number of streams increases in RSMA, the complexity of HARQ will increase. However, is it necessary to retransmit all the failed streams? For example, if $s_{1,1}$ cannot be decoded, is it necessary to retransmit all the streams, $s_{1,1}$, $s_2$ and $s_{1,2}$ in the retransmission rounds? In fact, we do not need always to do this. In some situations, a stream can be decoded once the previous one is decoded and cancelled. In the previous example, sometimes $s_2$ can be decoded directly after decoding $s_{1,1}$, so we do not need to retransmit $s_2$. Not only do the failed streams not always be retransmitted together, but neither do $s_{1,1}$ and $s_{1,2}$ need to be retransmitted at the same time. Let $r_{1,1}$ and $r_{1,2}$ denote the rate of $s_{1,1}$ and $s_{1,2}$, respectively. For a given rate requirement $r_1$, if $r_{1,1}+r_{1,2}\geq r_1$ holds, the message from user 1 can be decoded. In other words, if $r_{1,1}\geq r_1-r_{1,2}$ holds, this message can be decoded. Therefore, we can only retransmit $s_{1,1}$, because it is always possible to increase $r_{1,1}$ by CC or IR during the retransmission turns and fulfils $r_{1,1}\geq r_1-r_{1,2}$, and the maximum achievable $r_{1,2}$ is related to $\alpha$. Hence, in this RSMA retransmission scheme, only $s_{1,1}$ and $s_2$ will be involved in retransmission. In the retransmission turn, the power allocation $\alpha$ will not be changed, i.e., the retransmission power of $s_{1,1}$ and $s_{1,2}$ will still be $\alpha$ and $1-\alpha$, respectively. This can decrease the complexity at the receiver, otherwise, the receiver should be always informed about the new value of $\alpha$. Thus, $\alpha$ does not need to be decided at every turn, and it can be decided when both user 1 and user 2 send new packets. 

In this scheme, only $s_{1,1}$ and $s_2$ are involved in retransmission, and sometimes they do not need to be retransmitted together if they both fail. Channel conditions and power allocation can decide which streams will be retransmitted. Let us first consider in which situation the streams do not need to be retransmitted, and then the opposite situations would be retransmissions are needed. Let $g_1$ and $g_2$ denote the instantaneous channel gain of user 1 and user 2 in the current round, respectively. The SINR of $s_{1,1}$ is
\begin{equation}\label{gamma_{1,1}}
    \sigma_{1,1}=\frac{\alpha g_1 }{1+(1-\alpha) g_1 + g_2},
\end{equation}
the SINR of $s_2$ is 
\begin{equation}\label{gamma_2}
    \sigma_2=\frac{g_2}{1+(1-\alpha)g_1},
\end{equation}
and the SINR of $s_{1,2}$ is
\begin{equation}\label{gamma_{1,2}}
    \sigma_{1,2}=(1-\alpha)g_1.
\end{equation}
Obviously, if there is any $\alpha$ that can let
\begin{equation}\label{situation1}
\log_2(1+\sigma_{1,1})+\log_2(1+\sigma_{1,2}) \geq r_1,
\end{equation}
and 
\begin{equation}\label{situation2}
\log_2(1+\sigma_2)\geq r_2,
\end{equation}
both hold, the three streams do not require any retransmission. From (\ref{situation1}), we can obtain
\begin{equation}
    \alpha \leq \frac{(2^{r_1}-1)(1+g_1+g_2)-g_1-g_1 g_2-g_1^2}{g_1(2^{r_1}-1-g_1-g_2)} = \alpha_h,
\end{equation}
and from (\ref{situation2}) we can obtain
\begin{equation}
\alpha\geq 1+\frac{1}{g_1}-\frac{g_2}{g_1(2^{r_2}-1)} = \alpha_l.
\end{equation}
$\alpha_h$ and $\alpha_l$ may not be values between $0$ and $1$, and only when they are smaller than $1+\frac{1}{g_1}$, they are mathematically meaningful, because $1+\sigma_{1,1}$, $1+\sigma_{1,2}$ and $1+\sigma_2$ should be positive. Thus, the $\alpha$ which can let all the streams be decoded should satisfy
\begin{equation}\label{situation_of_decoding}
\alpha=\forall x\in S, \ S=\left\{y: 0\leq y \leq 1, \alpha_l\leq y \leq \alpha_h,\alpha_h\geq\alpha_l \right\},
\end{equation}
which means $\alpha$ can be any values in a set $S$ and $S$ is composed of the numbers between $\max \left(0,\alpha_l \right)$ and $\min \left(1,\alpha_h \right)$, as long as $\alpha_h \geq \alpha_l$.

\begin{figure}[]
\centering
\hspace{-2cm}\includegraphics[width=0.55\textwidth]{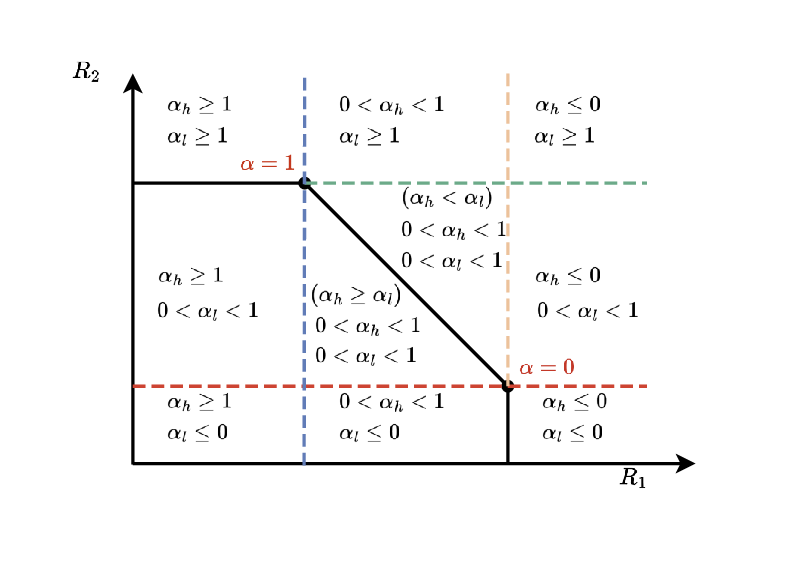}

\caption{An illustration of capacity region for two users.}\label{fig:rate_region}
\end{figure}

Although $\alpha_h$ and $\alpha_l$ may not be values between 0 and 1, they can still indicate whether the desired rate pairs $(r_1,r_2)$ are in the capacity region. Fig. \ref{fig:rate_region} is an illustration representing the relation between the values of $\alpha_h$ and $\alpha_l$ and the location of $(r_1,r_2)$. The top left corner point represents $\alpha=1$ because this point is achieved by allocating all the power to $s_{1,1}$ and decoding $s_{1,1}$ before $s_2$. The bottom right corner point represents $\alpha=0$ since this point is achieved by allocating all the power to $s_{1,2}$ and decoding $s_2$ before $s_{1,2}$. $\alpha$ between 0 and 1 indicates the points on the diagonal line. When $\alpha_h \leq 0$, the $\alpha$ satisfying (\ref{situation1}) does not exist, so it represents the points on the right side of the yellow line, which means this $r_1$ cannot be achieved regardless of $\alpha$. $\alpha_h\geq1$ means $\alpha$ always satisfies (\ref{situation1}), and this represents $r_1$ is always achievable and the corresponding points are located on the left side of the blue line. $0<\alpha_h<1$ represents the points between the blue line and the yellow line since $r_1$ is not always achievable. Similarly, $\alpha_l \leq 0$ represents the points below the red line, since any $\alpha$ can satisfy (\ref{situation2}). While $\alpha_l \geq 1$ represents the points above the green line because no $\alpha$ can satisfy (\ref{situation2}), and $0<\alpha_l<1$ represents the points between the red line and the green line.

Above analysis presents the relation between $\alpha_h$, $\alpha_l$ and the desired rate pairs $(r_1,r_2)$. If an $\alpha$ satisfying (\ref{situation_of_decoding}) exists, the desired rate pair is inside the capacity region, as shown in Fig. \ref{fig:rate_region}, and no retransmission is needed. However, for some values of $\alpha_h$ and $\alpha_l$, $S$ in (\ref{situation_of_decoding}) is empty, which means $\alpha$ satisfying (\ref{situation_of_decoding}) does not exist and the rate pair is outside the capacity region. In the following situations, $S$ is empty and retransmissions are needed.

\begin{itemize}
\item [1)] 
$\alpha_h \geq 1$ and $\alpha_l \geq 1$: any $\alpha$ can satisfy (\ref{situation1}); while no $\alpha$ can satisfy (\ref{situation2}). In other words, $s_2$ needs to be retransmitted irrespective of $\alpha$, while $s_{1,1}$ and $s_{1,2}$ do not need retransmission, so only $s_2$ needs to be retransmitted. 

\item[2)] $0<\alpha_h<1$ and $\alpha_l\geq1$: if $\alpha\leq\alpha_h$, it can satisfy (\ref{situation1}); while no $\alpha$ can satisfy (\ref{situation2}). Therefore, if we choose $\alpha \in [0,\alpha_h]$, only $s_2$ needs to be retransmitted. While if we choose $\alpha \in (\alpha_h,1]$, both $s_{1,1}$ and $s_2$ should be retransmitted, because this $\alpha$ does not satisfy (\ref{situation1}) and (\ref{situation2}).

\item[3)] $\alpha_h \geq 0$ and $\alpha_l\geq1$: no $\alpha$ can satisfy neither (\ref{situation1}) nor (\ref{situation2}), so both $s_{1,1}$ and $s_2$ need retransmission.

\item[4)] $0<\alpha_h<1$ and $0<\alpha_l<1$ $(\alpha_h<\alpha_l)$: (\ref{situation1}) and (\ref{situation2}) cannot be satisfied at the same time. Thus, if we choose $\alpha\in[0,\alpha_h]$, only $s_2$ is retransmitted. While if $\alpha \in [\alpha_l,1]$, only $s_{1,1}$ needs to be retransmitted because $s_2$ will be decoded after decoding $s_{1,1}$. Otherwise, $\alpha\in(\alpha_h,\alpha_l)$, both $s_{1,1}$ and $s_2$ are retransmitted.

\item[5)] $\alpha_h \leq 0$ and $0<\alpha_l<1$: There is $\alpha$ satisfying (\ref{situation2}) if $\alpha\geq\alpha_l$, but no $\alpha$ satisfied (\ref{situation1}). So if $\alpha \in [\alpha_l,1]$, only $s_{1,1}$ needs to be retransmitted; while if $\alpha \in [0,\alpha_l)$, both $s_{1,1}$ and $s_2$ will be retransmitted.

\item[6)]: $\alpha_h\leq0$ and $\alpha_l\leq 0$: Any $\alpha$ can satisfy (\ref{situation2}) but no $\alpha$ satisfies (\ref{situation1}). Thus, only $s_{1,1}$ will be retransmitted because $s_2$ will be decoded once $s_{1,1}$ is decoded. 

\end{itemize}
These situations are also summarized in TABLE \ref{tab:1}.

Apparently, for the situations that need retransmissions, the choice of $\alpha$ can decide which stream(s) will be retransmitted and the error probabilities after retransmissions. Since the two users operate the same service, their reliability requirements will be the same. Let $p_{1,1}$ and $p_2$ denote the error probability of $s_{1,1}$ and $s_2$ after the next retransmission turn, respectively. The $\alpha$ which gives the lowest $p_{1,1}+p_2$ will be chosen. Although sometimes it is possible to only retransmit one stream, considering all the possibilities is for reliability reasons. When $\alpha_h\leq0$ and $0<\alpha_l<1$, if only $s_{1,1}$ is retransmitted and it fails again after retransmission, $s_2$ still cannot be decoded; while if both streams are retransmitted, even if $s_1$ fails, $s_2$ still has some chance to be decoded. The sum of error probabilities of the former situation could be higher than the latter one. $p_{1,1}$ and $p_2$ in CC and IR have some differences, and the error probabilities of CC and IR will be analyzed separately.

\begin{table}[]
\renewcommand{\arraystretch}{1.5}
\caption{Choosing $\alpha$ according to $\alpha_h$ and $\alpha_l$}
\label{tab:1}
\centering
\begin{tabular}{|*{4}{c|}}
\hline
\diagbox{$\alpha_l$}{$\alpha_h$} &$\alpha_h\geq1$ & $0<\alpha_h<1$ & $\alpha_h \leq0$ \\
\hline

$\alpha_l\geq1$ & \thead{$s_2$ is retrans-\\mitted and \\ $\alpha\in[0,1]$.}  & \thead{1) $ \alpha\in[0,\alpha_h]$:  $s_2$\\ is retransmitted.\\2) $\alpha\in(\alpha_h,1]$:\\ both streams are \\retransmitted } &  \thead{$s_{1,1}$ and $s_2$ \\are  retrans-\\mitted, and \\ $\alpha\in[0,1]$.}  \\
\hline

$0<\alpha_l<1$ &
 $\forall \alpha \in [\alpha_l,1]$
 
 & \thead{1) $\alpha_h\geq\alpha_l$: \\$\forall \alpha \in [\alpha_l,\alpha_h]$.
 \\2) $\alpha_h<\alpha_l$: \\if $\alpha\in [\alpha_h,1]$, $s_{1,1}$\\ is retransmitted;\\ if $\alpha\in[0,\alpha_l]$, $s_2$ \\is retransmitted;\\ if $\alpha\in(\alpha_h,\alpha_l)$,\\ $s_{1,1}$ and $s_2$ are\\ retransmitted.} 
 
 & \thead{1) if $\alpha\in[\alpha_l,1]$, \\$s_{1,1}$ is\\ retransmitted. \\ 
2) if $\alpha\in[0,\alpha_l]$, \\$s_{1,1}$ and $s_2$ \\are retrans-\\mitted.} \\

\hline

$\alpha_l\leq0$ 
& \thead{$\alpha=0$ or \\ $\alpha=1$.}
& $\forall \alpha\in[0,\alpha_h]$
&   \thead{$s_{1,1}$ is retra-\\nsmitted, and \\$\alpha\in[0,1]$.} 
\\
\hline
\end{tabular}
\end{table}

\begin{figure}   
\centering
     \begin{subfigure}[b]{0.48\textwidth}         
         \includegraphics[scale=0.5]{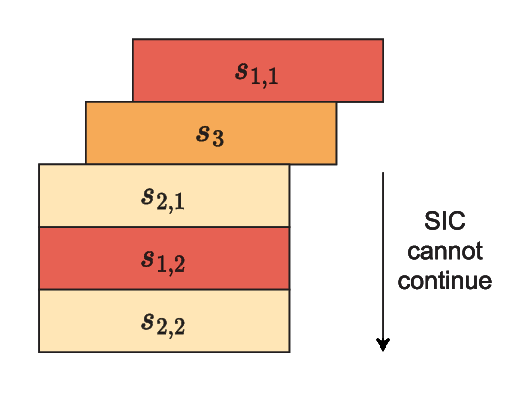}
         \caption{Splitting into two parts}
         \label{fig:two}
     \end{subfigure}     
     \begin{subfigure}[b]{0.48\textwidth}
         \includegraphics[scale=0.5]{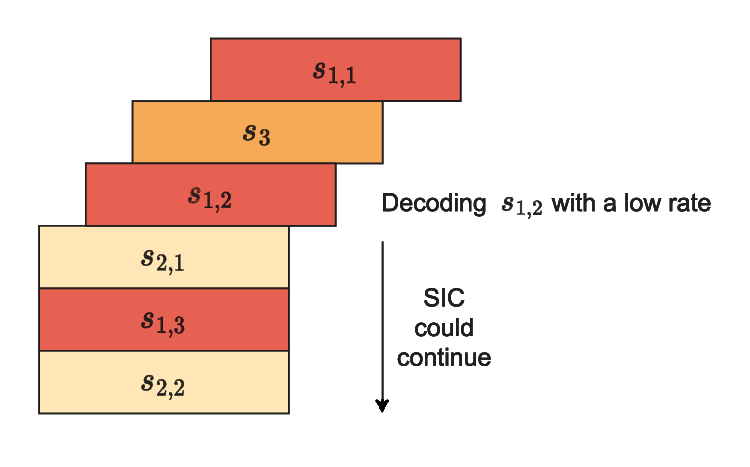}
         \caption{Splitting into three parts}
         \label{fig:three}
     \end{subfigure}   
    \caption{A 3-user case}
    \label{fig:A 3-user case}
\end{figure}

\begin{figure}   
\centering
     \begin{subfigure}[b]{0.48\textwidth}         
         \includegraphics[scale=0.5]{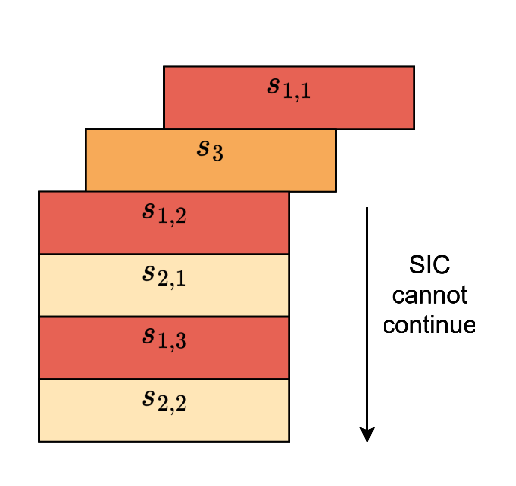}
         \caption{Splitting into two parts}
         \label{fig:two_split}
     \end{subfigure}     
     \begin{subfigure}[b]{0.48\textwidth}
         \includegraphics[scale=0.5]{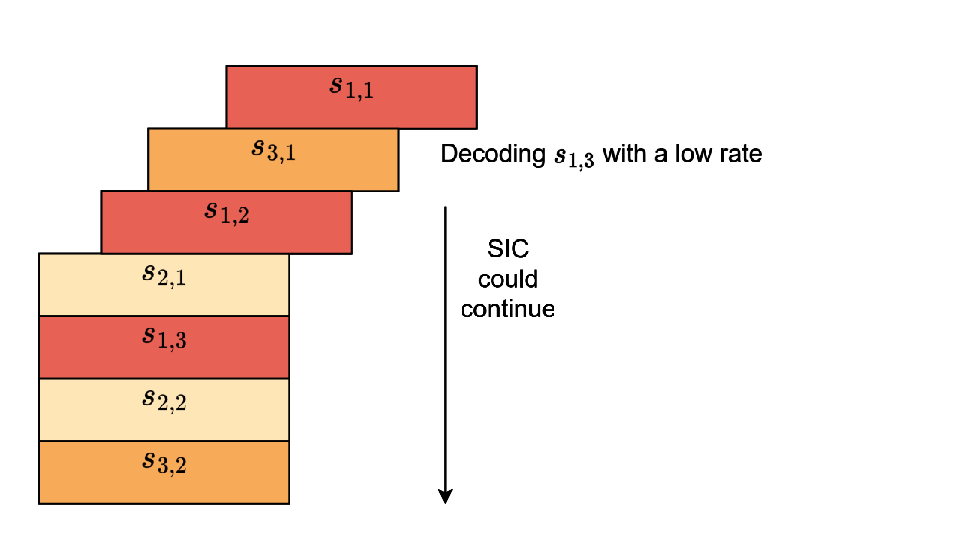}
         \caption{Splitting into three parts}
         \label{fig:three_split}
     \end{subfigure}   
    \caption{The number of users splitting messages.}
    \label{fig:split}
\end{figure}

\begin{remark}
Although this work focuses on a two-user case, it can provide some insights for a general $K$-user case. We first discuss the most intuitive design. In a $K$-user case, the capacity region can be achieved by splitting $K-1$ users \cite{485709}. We assume that $k\in \mathcal{K}, \mathcal{K}=\{1,2,...K\}$ and user $K$ does not split its message and it transmits $s_K$, other users split their messages into $s_{k,1}$ and $s_{k,2}$, where $k\in \mathcal{K}\setminus K$. Similar to the two-user case, we do not need to retransmit $s_{k,2}$ because the rate requirement can be fulfilled by retransmitting $s_{k,1}$. Therefore, a proper power allocation strategy would be important.

To handle the collisions better, a user may split its message into more than two parts to have more flexibility. A simple 3-user case example can give some intuitions. For a 3-user case, the boundary points of the capacity region can be reached by splitting the messages of any two users. We first consider that user 1 and user 2 split their message into two parts, and the BS receives the streams $s_{1,1}$, $s_{1,2}$, $s_{2,1}$, $s_{2,2}$ and $s_3$. We assume that $s_{1,1}$ and $s_3$ can be decoded, and $s_{1,2}$, $s_{2,1}$ and $s_{2,2}$ remain, which is shown in Fig. \ref{fig:two}. In this case, $s_{2,1}$ needs to be retransmitted. If user 1 splits its message into three parts, and the power fraction of $s_{1,1}$ maintains the same, the BS will receive $s_{1,1}$, $s_{1,2}$, $s_{1,3}$, $s_{2,1}$, $s_{2,2}$ and $s_3$. Then $s_{1,1}$ and $s_3$ are decoded, and $s_{1,2}$, $s_{1,3}$, $s_{2,1}$ and $s_{2,2}$ remain. However, $s_{1,2}$ could be decoded with a low rate, and the BS may continue decoding the remaining streams, which is shown in Fig. \ref{fig:three}. If BS still cannot continue decoding, $s_{2,1}$ will be retransmitted. This toy example shows that splitting a message into multiple parts would help handle collisions. 

Splitting all the messages may also be helpful in HARQ context and an illustration is shown in Fig. \ref{fig:split}. $s_3$ may not be decoded successfully and the SIC procedure stops, which is shown in Fig. \ref{fig:two_split}. If user 3 also splits its message into $s_{3,1}$ and $s_{3,2}$, $s_{3,1}$ can be decoded at a low rate and then the SIC could continue. This could handle collision better, but it will introduce more complexity. Although all users split messages may be beneficial for a HARQ scheme of RSMA, we note that it is not necessary to split all the messages to achieve the capacity region \cite{485709}.

The above two toy examples show that in a general $K$-user case, splitting a message into multiple parts and splitting all messages could be helpful for the design of HARQ for RSMA. However, when the number of steams increases, the power allocation and decoding order should be considered carefully. They are interesting topics and could be left for future work.
\end{remark}

\subsection{Chase Combining}\label{sec_2_2}
When CC is used, BS can use both previous unsuccessful copies and the new packet to decode the message, and the decoding error probability after $l$ retransmissions is given by \cite{930931},
\begin{equation}
    p_{error}^l=\mathrm{Pr} \left\{ \log_2 \left(  1+\sum_{i=0}^l \mathrm{SINR}_i \right) <  R\right\},
\end{equation}
where $\mathrm{SINR}_i$ is the SINR of $i$th retransmission round, and $R$ is the desired rate. For RSMA, finding the error probability with consideration of $l$ times retransmissions can be extremely complicated, so the choice of $\alpha$ only depends on the current round and the very next retransmission round. Note that we first consider the general $0<\alpha<1$, and $\alpha=0$ and $\alpha=1$ are two special cases and they will be discussed later.

First, let us consider $p_{1,1}$ when only $s_{1,1}$ is retransmitted, which is shown in Fig. \ref{fig:RSMA}. If $s_{1,1}$ fails, $s_{1,1}$, $s_2$ and $s_{1,2}$ will all receive NACK, but only $s_{1,1}$ is retransmitted. After successfully decoding $s_{1,1}$, then $s_2$ and $s_{1,2}$ will be decoded, so $p_2$ is exactly $p_{1,1}$. Since we have obtained instantaneous channel gains $g_1$ and $g_2$, in the following parts $h_1$ and $h_2$ denote the channels in the retransmission turn and they are all variables. BS uses the previous interfered copy of $s_{1,1}$ and retransmited $s_{1,1}$ to jointly decode $s_{1,1}$. $p_{1,1}$ can be represented as 
\begin{equation}
    p_{1,1}=\mathrm{Pr}\left\{\log_2(1+\sigma_{1,1}+\alpha|h_1|^2)<r_1-\log_2(1+\sigma_{1,2})\right\},
\end{equation}
and it can be rewritten as
\begin{equation}
    p_{1,1}=\mathrm{Pr}\left\{|h_1|^2<\frac{2^{r_1}}{\alpha\left(1+\sigma_{1,1}\right)}-\frac{1+\sigma_{1,1}}{\alpha}=\gamma_{cc_{1,1}}\right\},
\end{equation}
where $\gamma_{cc_{1,1}}$ is a term denoting 'residual SNR' for $s_{1,1}$. Introducing this term is enlightened by \cite{9254087}, and it represents the needed signal power to decode the stream. Since CC and IR are discussed in separate sections, the subscript of HARQ type is omitted for simplicity, so $\gamma_{cc_{1,1}}$ and $\gamma_{ir_{1,1}}$ will be presented as $\gamma_{1,1}$ in this subsection and next subsection, respectively. The subscripts of HARQ type will also be omitted in other variables in the same way. The distribution of the channel gain is exponential, and the probability density function (pdf) is 
\begin{equation}\label{channel_gain_pdf}
f \left( x,\Gamma_k \right)=\frac{1}{\Gamma_k}e^{-\frac{x}{\Gamma_k}},
\end{equation}
so that
\begin{equation}\label{p_11_1}
    p_{1,1} =1-e^{-\frac{\gamma_{1,1}}{\Gamma_1}},
\end{equation}
and $p_{1,1}$ increases as $\gamma_{1,1}$ decreases. The $\alpha$ which gives the lowest $p_{1,1}$ can be found by sequential quadratic programming (SQP).

\begin{figure}[]
\centering
\includegraphics[width=0.5\textwidth]{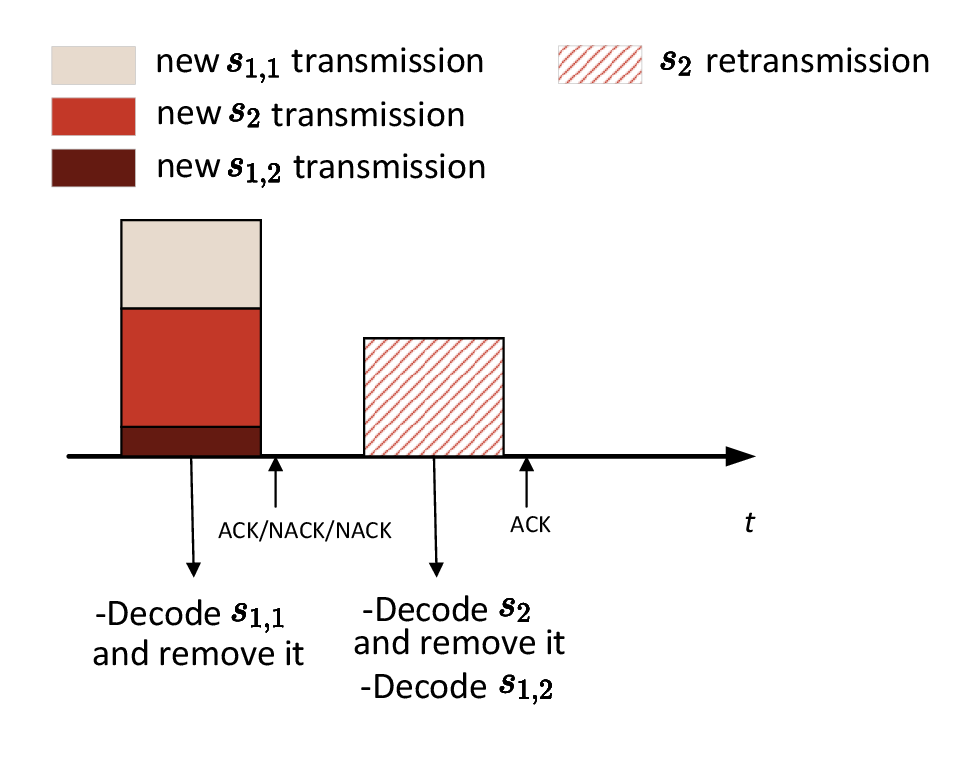}
\caption{The situation that only $s_2$ needs retransmission. BS fails to decode $s_2$ and $s_{1,2}$, but only asks user 2 to retransmit $s_2$. Then, BS uses two interfered copies of $s_2$ to decode it, and then continues decoding $s_{1,2}$. }\label{fig:retransmit_s2}
\end{figure}

Then, $p_2$ when only $s_2$ is retransmitted is analyzed and an illustration is shown in Fig. \ref{fig:retransmit_s2}. The three streams are sent to BS. BS decodes and cancels $s_{1,1}$ first, but it fails to decode $s_2$, so $s_{1,2}$ also cannot be decoded. BS only asks user 2 to retransmit $s_2$. Then, BS can use the unsuccessful copies of $s_2$ with interference and the retransmitted $s_2$ to decode $s_2$ and cancel it, and then continue decoding $s_{1,2}$. Thus,
\begin{equation}
\begin{aligned}
    p_{2} & = \mathrm{Pr}\left\{ |h_2|^2<2^{r_2}-1-\sigma_{2}= \gamma_{2} \right\} \\ &=1-e^{-\frac{\gamma_2}{\Gamma_2}},
\end{aligned}
\end{equation}
where $\gamma_2$ is the residual SNR for $s_2$. Similar to $p_{1,1}$, $p_2$ decreases as $\gamma_2$ decreases. Obviously, the highest applicable $\alpha$ brings the lowest $\gamma_2$, so for this situation the optimal $\alpha=\alpha_h$.

\begin{figure}[]
\centering
\includegraphics[width=0.5\textwidth]{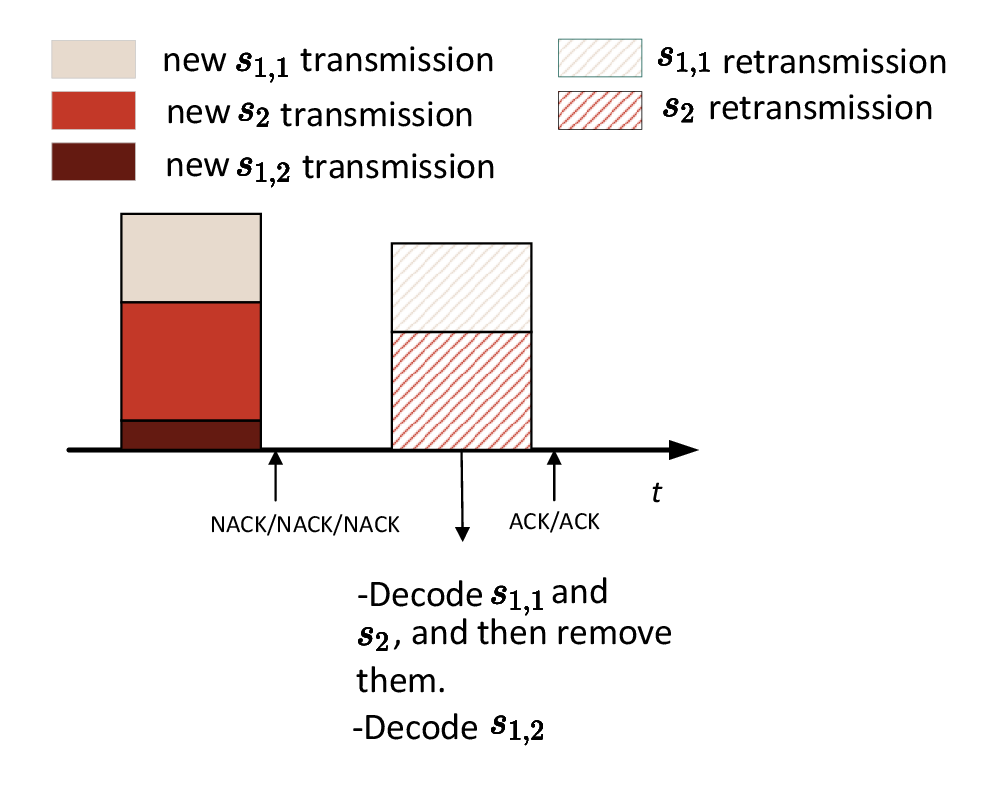}
\caption{The situation that both streams need retransmission. BS fails to decode $s_{1,1}$ and asks both user 1 and user 2 to retransmit. Then, BS uses two interfered copies of $s_{1,1}$ and $s_2$ to decode them and then continues decoding $s_{1,2}$. }\label{fig:retransmit_s11_s2}
\end{figure}
Finally, $p_{1,1}$ and $p_2$ when both $s_{1,1}$ and $s_2$ need to be retransmitted are analyzed, and this is shown in Fig. \ref{fig:retransmit_s11_s2}. Although we assume that it is always possible to find a low rate to decode $s_{1,2}$, the prerequisite is that $s_2$ has been decoded. In the previous two situations, once $s_{1,1}$ or $s_2$ is decoded, the decoding process can continue, but here $s_{1,1}$ and $s_2$ do not depend on each other completely, so here actually $p_{1,1}$ is considered as the error probability of whole message from user 1. Thus,
\begin{equation}
\begin{aligned}\label{p11_origin}
p_{1,1}&=\mathrm{Pr}\left\{\frac{\alpha|h_1|^2}{1+|h_2|^2}<\gamma_{1,1}^{(1)}, \frac{|h_2|^2}{1+\alpha|h_1|^2}<\gamma_2^{(1)}\right\}\\
&+\mathrm{Pr}\left\{\frac{|h_2|^2}{1+\alpha|h_1|^2}\geq\gamma_{2}^{(1)}, \alpha|h_1|^2<\gamma_{1,1}^{(2)} \right\}\\
&+\mathrm{Pr}\left\{\frac{\alpha|h_1|^2}{1+|h_2|^2}\geq\gamma_{1,1}^{(1)}, |h_2|^2<\gamma_2^{(2)}\right\},
\end{aligned}
\end{equation}
where
\begin{equation}
    \gamma_{1,1}^{(1)}=\frac{2^{r_1}}{1+\sigma_{1,2}}-1-\sigma_{1,1},
\end{equation}
\begin{equation}
    \gamma_{2}^{(1)}=2^{r_2}-1-\frac{g_2}{1+g_1},
\end{equation}
\begin{equation}
    \gamma_{1,1}^{(2)}=\frac{2^{r_1}}{1+\sigma_{1,2}}-1-\frac{\alpha g_1}{1+(1-\alpha)g_1},
\end{equation}
and
\begin{equation}
    \gamma_{2}^{(2)}=2^{r_2}-1-\sigma_2.
\end{equation}
The error probability of $s_2$ can be represented as
\begin{equation}\label{p2_origin}
\begin{aligned}
    p_{2}= & \mathrm{Pr}\left\{\frac{\alpha|h_1|^2}{1+|h_2|^2} <\gamma_{1,1}^{(1)}, 
      \frac{|h_2|^2}{1+\alpha|h_1|^2}< \gamma_{2}^{(1)}
    \right\} \\
    & +\mathrm{Pr}\left\{\frac{\alpha|h_1|^2}{1+|h_2|^2}\geq \gamma_{1,1}^{(1)}, |h_2|^2<\gamma_{2}^{(2)}
    \right\}.
\end{aligned}
\end{equation}
The derivation of $p_{1,1}$ and $p_2$ is in Appendix \ref{FirstAppendix}, and the results are (\ref{p_11_s11_s2}) and (\ref{p_2_s11_s2}), respectively,
\begin{figure*}[]
\normalsize
\begin{equation}
\label{p_11_s11_s2}
 p_{1,1}=
    \begin{cases}
      \underbrace{ 1-\frac{\Gamma_1\alpha}{\Gamma_2 \gamma_{1,1}^{(1)}+\Gamma_1\alpha}e^{-\frac{\gamma_{1,1}^{(1)}}{\Gamma_1\alpha}}-\frac{\Gamma_2}{\Gamma_2+\Gamma_1\gamma_{2}^{(1)}\alpha}e^{-\frac{\gamma_{1,1}^{(2)}}{\Gamma_1\alpha}-\frac{\gamma_{2}^{(1)}\gamma_{1,1}^{(2)}}{\Gamma_2}-\frac{\gamma_{2}^{(1)}}{\Gamma_2}}+\frac{\Gamma_1}{\Gamma_1+\Gamma_2\gamma_{1,1}^{(1)}}e^{-\frac{\gamma_{1,1}^{(1)}}{\Gamma_1}}\left[1-e^{-\frac{\gamma_{2}^{(2)}}{\Gamma_2}-\frac{\gamma_{1,1}^{(1)}\gamma_{2}^{(2)}}{\Gamma_1}} \right]}_A, & \mathrm{if}\ \gamma_{1,1}^{(1)} \gamma_{2}^{(1)}\geq 1 \\
      A-\frac{\Gamma_2\gamma_{1,1}^{(1)}}{\Gamma_2\gamma_{1,1}^{(1)}+\Gamma_1\alpha}e^{\frac{1}{\Gamma_2}-\left(\frac{1}{\Gamma_1}+\frac{\alpha}{\Gamma_2\gamma_{1,1}^{(1)}} \right)c}+\frac{\Gamma_2}{\Gamma_2+\Gamma_1\gamma_{2}^{(1)}\alpha}e^{-\frac{\gamma_{2}^{(1)}}{\Gamma_2}-\left(\frac{1}{\Gamma_1}+\frac{\gamma_{2}^{(1)}\alpha}{\Gamma_2}\right)c}, & \text{otherwise}
    \end{cases}
\end{equation}
\begin{equation}\label{p_2_s11_s2}
     p_{2}=
    \begin{cases}
      \underbrace{1-\frac{\Gamma_1\alpha}{\Gamma_2\gamma_{1,1}^{(1)}+\Gamma_1\alpha}e^{-\frac{\gamma_{1,1}^{(1)}}{\Gamma_1\alpha}}-\frac{\Gamma_2}{\Gamma_2+\Gamma_1\gamma_{2}^{(1)}\alpha}e^{-\frac{\gamma_{2}^{(1)}}{\Gamma_2}}+\frac{\Gamma_1}{\Gamma_1+\Gamma_2\gamma_{1,1}^{(1)}}e^{-\frac{\gamma_{1,1}^{(1)}}{\Gamma_1}}\left[1-e^{-\frac{\gamma_{2}^{(2)}}{\Gamma_2}-\frac{\gamma_{1,1}^{(1)}\gamma_{2}^{(2)}}{\Gamma_1}} \right]}_B, & \mathrm{if}\ \gamma_{1,2}^1 \gamma_{2}^{(1)}\geq 1

      \\
      B-\frac{\Gamma_2\gamma_{1,1}^{(1)}}{\Gamma_2\gamma_{1,1}^{(1)}+\Gamma_1\alpha}e^{\frac{1}{\Gamma_2}-\left(\frac{1}{\Gamma_1}+\frac{\alpha}{\Gamma_2\gamma_{1,1}^{(1)}} \right)c}+\frac{\Gamma_2}{\Gamma_2+\Gamma_1\gamma_{2}^{(1)}\alpha}e^{-\frac{\gamma_{2}^{(1)}}{\Gamma_2}-\left(\frac{1}{\Gamma_1}+\frac{\gamma_{2}^{(1)}\alpha}{\Gamma_2}\right)c}, & \text{otherwise}
    \end{cases}
\end{equation}
\hrulefill
\vspace*{4pt}
\end{figure*} 
where $c=-\frac{\left(1+\gamma_{2}^{(1)}\right)\gamma_{1,1}^{(1)}}{\alpha\left(\gamma_{2}^{(1)}\gamma_{1,1}^{(1)}-1\right)}$. The $\alpha$ which brings the lowest $p_{1,1}+p_2$ will be chosen.

Although the above discussion is general for RSMA, $\alpha=0$ and $\alpha=1$ could be two special cases when considering the error probability. In the previous analysis, user 1 and user 2 will stop generating new packets in retransmission turns, because the failure of either $s_{1,1}$ or $s_2$ means that BS does not decode messages from user 1 and user 2 successfully. Similarly, if $\alpha=1$ and the message of user 1 denoted by $s_1$ fails, both user 1 and user 2 will pause transmitting new packets. For $\alpha=0$ and $s_2$ fails, the situation is the same. These two situations can be included in the above discussion. However, if $\alpha=1$ and $s_1$ can be decoded, user 1 will transmit a new packet in the next turn, so user 2 will retransmit $s_2$ along with a new $s_1$. Similarly, when $\alpha=0$ and $s_2$ can be decoded, $s_1$ will be retransmitted with a new $s_2$. In these situations, the error probability of the new message also needs to be considered, and this could affect the choice of $\alpha$. 

The formulations of error probabilities of these two situations are similar. We consider $\alpha=1$ first, and $s_2$ will be retransmitted with a new $s_1$. For user 1,
\begin{equation}\label{special_case_p_1}
\begin{aligned}
    p_{1}=&\mathrm{Pr}\left\{\frac{|h_1|^2}{1+|h_2|^2}<\gamma_{1}, \ \frac{|h_2|^2}{1+|h_1|^2}<\gamma_2\right\} \\ & +\mathrm{Pr}\left\{\frac{|h_2|^2}{1+|h_1|^2}\geq\gamma_2,\ |h_1|^2<\gamma_{1} \right\},
\end{aligned}
\end{equation}
and for user 2,
\begin{equation}\label{special_case_p_2}
\begin{aligned}
    p_2=& \mathrm{Pr}\left\{\frac{(|h_1|^2}{1+|h_2|^2}<\gamma_{1}, \ \frac{|h_2|^2}{1+|h_1|^2}<\gamma_2\right\} \\ & +\mathrm{Pr} \left\{ \frac{|h_1|^2}{1+|h_2|^2}\geq \gamma_{1},\ |h_2|^2<\gamma_2\right\},
\end{aligned}
\end{equation}
where
\begin{equation}
    \gamma_{1}=2^{r_1}-1,
\end{equation}
and
\begin{equation}
    \gamma_{2}=2^{r_2}-1-g_2.
\end{equation}
The results are (\ref{p_1}) and (\ref{p_2}), and the detailed derivation is in Appendix \ref{SecondAppendix}.

\begin{figure*}[]
\normalsize


\begin{equation}
\label{p_1}
 p_{1}=
    \begin{cases}
      \underbrace{1-\frac{\Gamma_1}{\gamma_1\Gamma_2+\Gamma_1}e^{-\frac{\gamma_1}{\Gamma_1}}-\frac{\Gamma_2}{\Gamma_2+\gamma_2\Gamma_1}e^{-\frac{\gamma_2}{\Gamma_2}-\frac{\gamma_1}{\Gamma_1}-\frac{\gamma_1\gamma_2}{\Gamma_2}}}_C, & \mathrm{if}\ \gamma_1\gamma_{2}\geq 1 \\
      C-\frac{\gamma_1\Gamma_2}{\gamma_1\Gamma_2+\Gamma_1}e^{\frac{1}{\Gamma_2}+\frac{\left(1+\gamma_2\right)\gamma_1}{\Gamma_1\left(\gamma_1\gamma_2-1\right)}+\frac{1+\gamma_2}{\Gamma_2\left(\gamma_1\gamma_2-1\right)}}+\frac{\Gamma_2}{\Gamma_2+\gamma_2\Gamma_1}e^{-\frac{\gamma_2}{\Gamma_2}+\frac{\gamma_1\left(1+\gamma_2\right)}{\Gamma_1\left(\gamma_1\gamma_2-1\right)}+\frac{\gamma_1\gamma_2\left(1+\gamma_2\right)}{\Gamma_2\left(\gamma_1\gamma_2-1\right)}}, & \text{otherwise}
    \end{cases}
\end{equation}

\begin{equation}
\label{p_2}
p_2=
\begin{cases}
 \underbrace{1-\frac{\Gamma_2}{\Gamma_2+\gamma_2\Gamma_1}e^{-\frac{\gamma_2}{\Gamma_2}}-\frac{\Gamma_1}{\Gamma_1+\gamma_1\Gamma_2}e^{-\frac{\gamma_1}{\Gamma_1}-\frac{\gamma_2}{\Gamma_2}-\frac{\gamma_1\gamma_2}{\Gamma_1}}}_D, & \mathrm{if}\ \gamma_1\gamma_{2}\geq 1 \\
      D-\frac{\gamma_1\Gamma_2}{\gamma_1\Gamma_2+\Gamma_1}e^{\frac{1}{\Gamma_2}+\frac{\left(1+\gamma_2\right)\gamma_1}{\Gamma_1\left(\gamma_1\gamma_2-1\right)}+\frac{1+\gamma_2}{\Gamma_2\left(\gamma_1\gamma_2-1\right)}}+\frac{\Gamma_2}{\Gamma_2+\gamma_2\Gamma_1}e^{-\frac{\gamma_2}{\Gamma_2}+\frac{\gamma_1\left(1+\gamma_2\right)}{\Gamma_1\left(\gamma_1\gamma_2-1\right)}+\frac{\gamma_1\gamma_2\left(1+\gamma_2\right)}{\Gamma_2\left(\gamma_1\gamma_2-1\right)}}, & \text{otherwise}
\end{cases}
\end{equation}
\vspace*{4pt}
\hrulefill
\end{figure*}

When $\alpha=0$, $s_1$ will be retransmitted with a new $s_2$. The expressions are as same as (\ref{special_case_p_1}) and (\ref{special_case_p_2}), and the difference is the values of $\gamma_1$ and $\gamma_2$, which are 
\begin{equation}\label{1}
    \gamma_1=2^{r_1}-1-g_1,
\end{equation}
and
\begin{equation}\label{2}
    \gamma_2=2^{r_2}-1,
\end{equation}
so by substituting (\ref{1}) and (\ref{2}) to (\ref{p_1}) and (\ref{p_2}), $p_1$ and $p_2$ can be obtained. Then the lowest $p_1+p_{2}$ can be obtained by SQP.

All the error probabilities of the situation in TABLE \ref{tab:1} can be obtained, and the $\alpha$ leads to the lowest error probabilities will be chosen. If the streams still cannot be decoded after one retransmission, the users will retransmit them again until the streams are decoded or they do not have more retransmission turns.

\subsection{Incremental Redundancy}\label{sec_2_3}
When IR is applied, the decoding error probability after $l$ retransmissions can be represented as
\begin{equation}
    p_{error}^l=\mathrm{Pr} \left\{ \sum_{i=0}^l\log_2 \left(  1+ \mathrm{SINR}_i \right) <  R\right\},
\end{equation}
and $\mathrm{SINR}_i$ is the SINR of the $i$th turn. $\alpha$ can be found by the similar method in Section \ref{sec_2_2}.

If only $s_{1,1}$ is retransmitted, which is presented in Fig. \ref{fig:RSMA}, $p_{1,1}$ is the same as (\ref{p_11_1}), but the 'residual SINR' has a difference. In IR,
\begin{equation}\label{sub_1}
    \gamma_{1,1}=\frac{2^{r_1}}{\alpha\left( 1+\sigma_{1,1}\right) \left(1+\sigma_{1,2} \right)}-\frac{1}{\alpha},
\end{equation}
and then impose (\ref{sub_1}) into (\ref{p_11_1}), and the $\alpha$ which gives the lowest $\gamma_{1,1}$ will be chosen.

For the situation where only $s_2$ is retransmitted shown in Fig. \ref{fig:retransmit_s2},
\begin{equation}
\begin{aligned}
    p_2=&\mathrm{Pr}\left\{|h_2|^2<\frac{2^{r_2}}{1+\sigma_{2}}-1=\gamma_2 \right\}\\
    &=1-e^{-\frac{\gamma_2}{\Gamma_2}}.
\end{aligned}
\end{equation}
$p_2$ decreases as $\alpha$ increases, so $\alpha_h$ gives the lowest $p_2$.

Then, when both $s_{1,1}$ and $s_{2}$ are retransmitted in Fig. \ref{fig:retransmit_s11_s2}, $p_{1,2}$ and $p_2$ can be represented by the same expressions in (\ref{p_11_s11_s2}) and (\ref{p_2_s11_s2}), respectively. But 
\begin{equation}\label{sub_2}
    \gamma_{1,1}^{(1)}=\frac{2^{r_1}}{\left(1+\sigma_{1,2} \right)\left(1+\sigma_{1,1}\right)}-1,
\end{equation}
\begin{equation}
    \gamma_{2}^{(1)}=\frac{2^{r_2}}{1+\frac{g_2}{1+g_1}}-1,
\end{equation}

\begin{equation}
    \gamma_{1,1}^{(2)}=\frac{2^{r_1}}{\left(1+\sigma_{1,2} \right) \left(1+\frac{\alpha g_1}{1+(1-\alpha)g_1}\right)}-1,
\end{equation}

and
\begin{equation}\label{sub_5}
    \gamma_{2}^{(2)}=\frac{2^{r_2}}{1+\sigma_2}-1.
\end{equation}
Substitute (\ref{sub_2})-(\ref{sub_5}) to (\ref{p_11_s11_s2}) and (\ref{p_2_s11_s2}), and then the $\alpha$ can be found.

For the special cases when $\alpha=0$ or $\alpha=1$, the expressions are also the same as in (\ref{special_case_p_1}) and (\ref{special_case_p_2}). When $\alpha=1$, $\gamma_1$ does not change, and 
\begin{equation}
    \gamma_2=\frac{2^{r_2}}{1+g_2}-1.
\end{equation}
When $\alpha=0$, $\gamma_2$ does not change, and 
\begin{equation}
    \gamma_1=\frac{2^{r_1}}{1+g_1}-1.
\end{equation}

\section{HARQ Scheme for FDMA and NOMA}\label{sec_2_4}
This section will introduce the HARQ scheme for FDMA and NOMA, and we use them as baselines and compare their performances with RSMA in Section \ref{sec:4}. 

In FDMA, users are allocated to non-interfering resources, so each user will be allocated to an isolated bandwidth. We assume that the bandwidth is normalized to 1, and each user is allocated a fraction of the bandwidth. If CC is used, the error probability of user $k$ after $L$ retransmissions is 
\begin{equation}
    p_k=\mathrm{Pr}\left\{w_k\log_2\left(1+\frac{\sum_{l=0}^L|h_k^{(l)}|^2}{w_k} \right)<r_k\right\},
\end{equation}
where $w_k$ is the fraction of bandwidth allocated to user $k$, $\sum_{k=1}^2 w_k=1$, $h_k^{(l)}$ is the channel of user k in $l$th round, and $r_k$ is the desired rate. The error probability when IR is used is
\begin{equation}
    p_k=\mathrm{Pr}\left\{\sum_{l=0}^L\left[w_k\log_2\left(1+\frac{|h_k^{(l)}|^2}{w_k} \right)\right]<r_k\right\},
\end{equation}
The $w_k$ which gives the lowest $\sum_{k=1}^2p_k$ will be chosen, and it can be computed by Monte Carlo simulation. Compared to RSMA, the retransmissions are independent in FDMA, i.e., whether a user will retransmit does not depend on other users, and the users retransmit whole streams.

NOMA can be seen as a subset of RSMA since it equals that $\alpha=0$ or $\alpha=1$. The decoding order is decided by the channel gain, which means if user 1 has better channel conditions, it will be decoded first ($\alpha=1$), otherwise, user 2 will be decoded first ($\alpha=0$). For CC, the error probability of user 1 is
\begin{equation}
    p_1=\mathrm{Pr}\left\{\log_2\left(1+\sum_{l=1}^L\mathrm{SINR}_1^{(l)} \right)<r_1\right\},
\end{equation}
where 
\begin{equation}\label{sinr_1}
\mathrm{SINR}_1^{(l)}=
    \begin{cases}
    \frac{|h_1|^2}{1+|h_2|^2},  &\mathrm{if} \ \alpha=1   \\
    |h_1|^2, &\mathrm{if} \ \alpha=0.
    \end{cases}
\end{equation}
The error probability of user 2 is
\begin{equation}
    p_2=\mathrm{Pr}\left\{\log_2\left(1+\sum_{l=1}^L \frac{|h_2^{(l)}|^2}{1+(1-\alpha)|h_1^{l}|^2} \right)<r_2\right\}.
\end{equation}
For IR, similarly, the error probability of user 1 is
\begin{equation}
    p_1=\mathrm{Pr}\left\{\sum_{l=1}^L\left[\log_2\left(1+\mathrm{SINR}_1^{(l)} \right)\right]<r_1\right\},
\end{equation}
where $\mathrm{SINR_1^{(l)}}$ is (\ref{sinr_1}), and error probability of user 2 is
\begin{equation}
    p_2=\mathrm{Pr}\left\{\sum_{l=1}^L\left[\log_2\left(1+ \frac{|h_2^{(l)}|^2}{1+(1-\alpha)|h_1^{l}|^2} \right)\right]<r_2\right\}.
\end{equation}
NOMA can be seen as a special case of RSMA, but it only retransmits whole streams, so it is not as flexible as RSMA.

\section{Numerical Results}\label{sec:4}
In this section, the error probabilities and transmission power per packet of RSMA with HARQ will be presented, and they are compared with FDMA and NOMA. For all the simulation results, the average channel gains, $\Gamma_1$ and $\Gamma_2$, are set to $20$ dB and $15$ dB, respectively. Since we assume that user 1 and user 2 require the same service, they have the same reliability requirement and the desired rate, and this rate is the x-axis in the figures.

\begin{figure}[]
\centering
\includegraphics[scale=0.5]{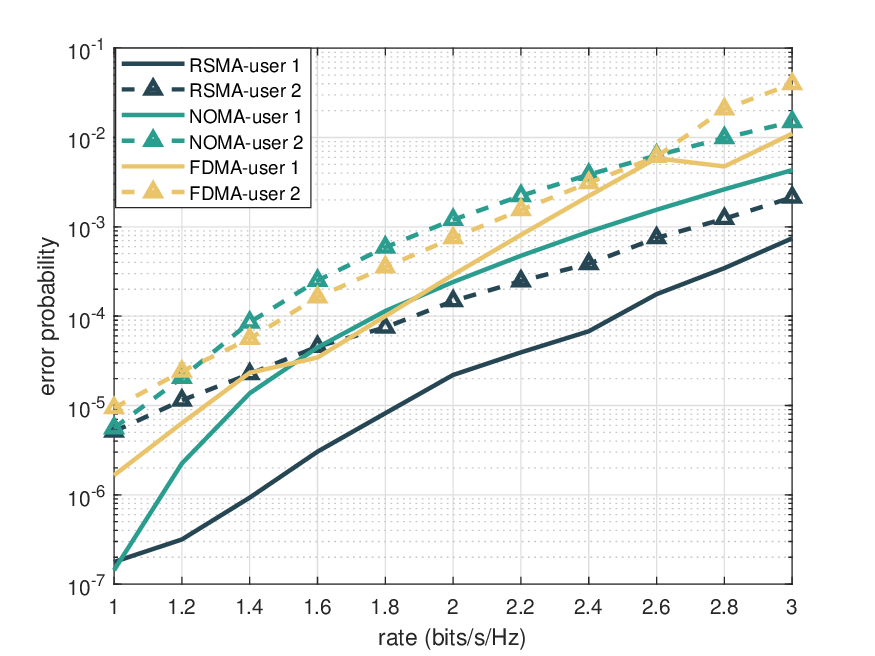}
\caption{Error probabilities of two users in CC. The average channel gains for user 1 and user 2 are $20$ dB and $15$ dB, respectively. Retransmission times $L=2$.}\label{fig:T_2_CC}
\end{figure}

\begin{figure}[]
\centering
\includegraphics[scale=0.5]{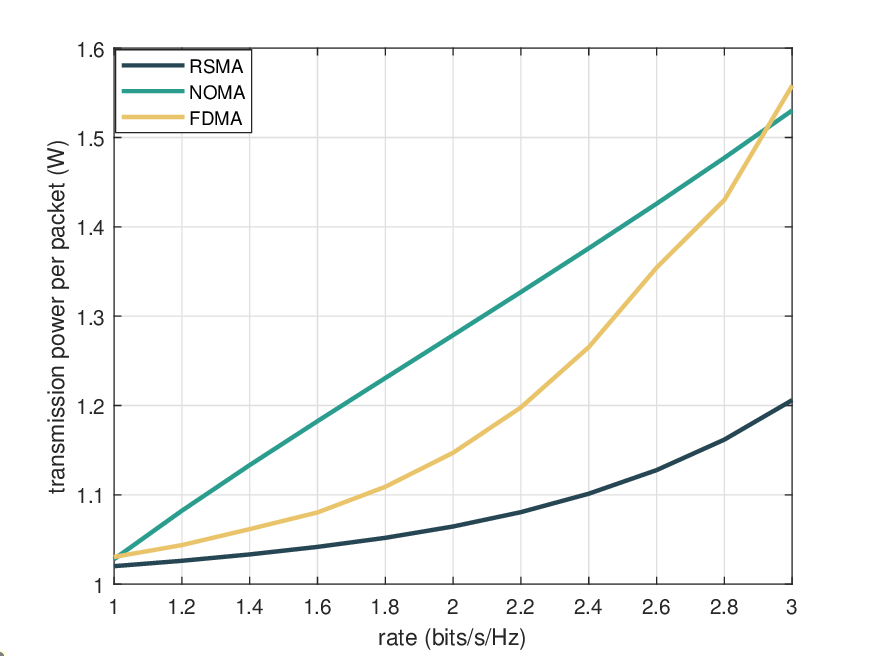}
\caption{Average power consumption of each packet in CC with 2 retransmissions.}\label{fig:T_2_CC_P}
\end{figure}

In Fig. \ref{fig:T_2_CC} and Fig. \ref{fig:T_2_CC_P}, two retransmissions are allowed with CC, i.e. for low latency service, URLLC. In Fig. \ref{fig:T_2_CC}, the error probabilities for different rates are presented, and they are obtained by Monte Carlo simulation. The solid lines denote user 1, and the dashed lines marked by triangles denote user 2. Generally, user 1 has a lower error probability than user 2 due to the better channel condition, and RSMA has the lowest error probability for user 2. Actually, the error probability of user 2 is more critical, because it also needs to meet the reliability requirement. That both users can have a low error probability is a main advantage of RSMA. The reason is that RSMA can exploit the resources better than FDMA, and it is more flexible than NOMA. In NOMA, if user 1 cannot be decoded, user 2 will also fail; while in RSMA, part of the interference of user 1 can be cancelled and user 2 still has the possibility to be decoded. Thus, RSMA has better performance for user 2. Fig. \ref{fig:T_2_CC_P} shows the average transmission power per packet. RSMA consumes the least energy. One reason is that it has the largest capacity region than FDMA and NOMA without time-sharing so it is more possible to achieve a given rate. The other reason is that when $s_{1,1}$ is retransmitted only $\alpha$ fraction of power is used. NOMA consumes more power than FDMA, because in NOMA $s_1$ is decoded with interference, so $s_1$ is retransmitted a bit more frequently than FDMA, which is free from interference from others. 

\begin{figure}[]
\centering
\includegraphics[scale=0.5]{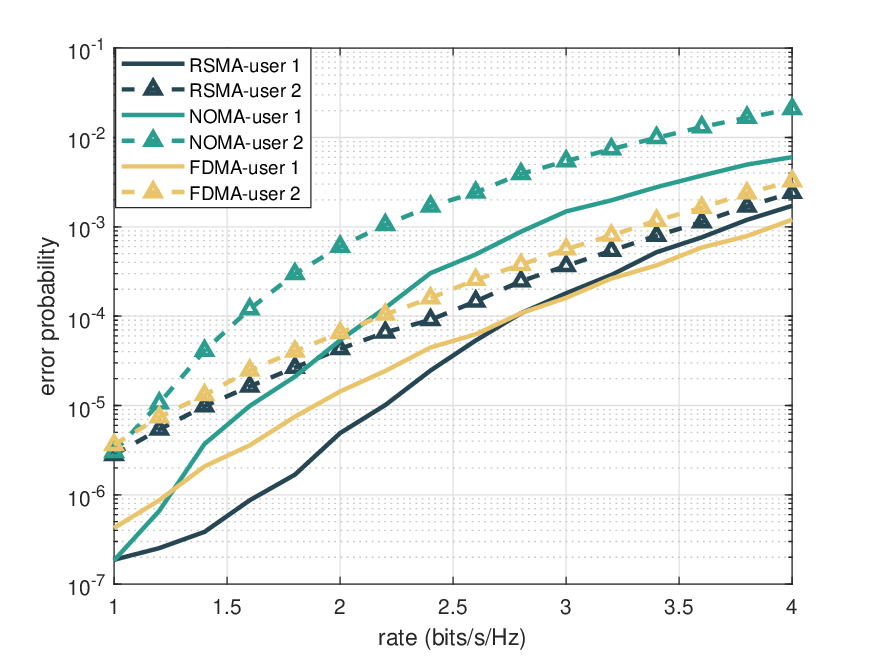}
\caption{Error probabilities of two users in IR. The average channel gains for user 1 and user 2 are $20$ dB and $15$ dB, respectively. Retransmission times $L=2$.}\label{fig:T_2_IR}
\end{figure}

\begin{figure}[]
\centering
\includegraphics[scale=0.5]{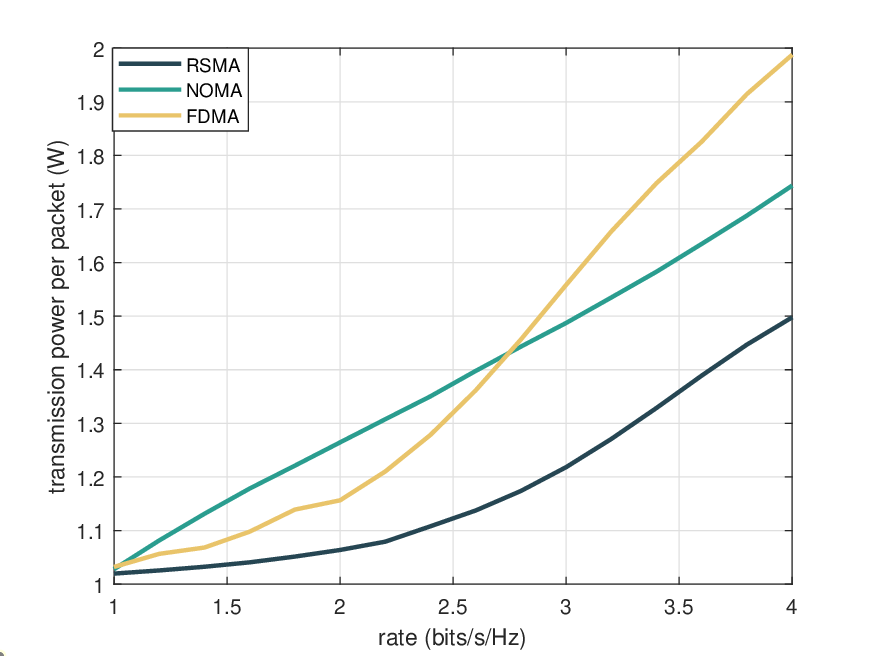}
\caption{Average power consumption of each packet in IR with 2 retransmissions.}\label{fig:T_2_IR_P}
\end{figure}

In Fig. \ref{fig:T_2_IR} and Fig. \ref{fig:T_2_IR_P}, IR is used and two retransmissions are allowed. Fig. \ref{fig:T_2_IR} presents the error probabilities of user 1 and user 2. RSMA with HARQ still has the best performance and the two users can all obtain low error probabilities. The trends of the curves are quite similar to the results in Fig. \ref{fig:T_2_CC}, but FDMA outperforms NOMA. FDMA benefits more from IR since the achievable rate is the sum of the rate in each turn, which mitigates the effect of limited bandwidth. While for NOMA the retransmission with interference may not contribute to the achievable rate as much as the non-interfering retransmission, so the error probability increases. However, since the decoding order of RSMA is shuffled, the negative impact of interference is mitigated to some extent. For example, in RSMA, when $\alpha$ is between 0 and 1 and either $s_{1,1}$ or $s_2$ needs to be retransmitted, the stream will be retransmitted alone without interference. But the decoding of $s_{1,2}$ depends on $s_2$, and this is the reason the error probability of user 1 is close to user 2 as the rate increases. While in NOMA, when $s_1$ is decoded before $s_2$, if $s_1$ can be decoded and $s_2$ fails, $s_2$ will be retransmitted with another new $s_1$, and $s_2$ could fail again due to the interference of this new stream. Fig. \ref{fig:T_2_IR_P} shows the average power consumption of each packet. Similar to the trends in Fig. \ref{fig:T_2_CC_P}, RSMA consumes the least power. NOMA consumes more power than FDMA when the rate is lower than around $2.7\ \mathrm{(bit/s/Hz)}$, because NOMA could need more retransmissions than FDMA. However, NOMA consumes less power when the rate is higher than $2.7\ \mathrm{(bit/s/Hz)}$ since FDMA is limited by isolated resources. 

\begin{figure}
\centering
\includegraphics[scale=0.5]{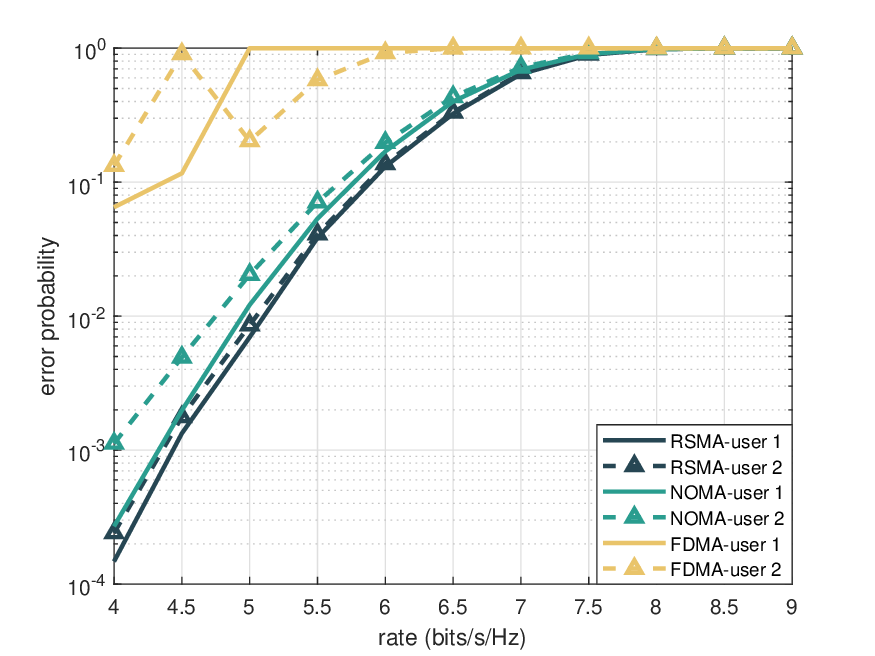}
\caption{Error probabilities of two users in CC. The average channel gains for user 1 and user 2 are $20$ dB and $15$ dB, respectively. Retransmission times $L=4$.}\label{fig:T_4_CC}
\end{figure}

\begin{figure}[]
\centering
    \includegraphics[scale=0.5]{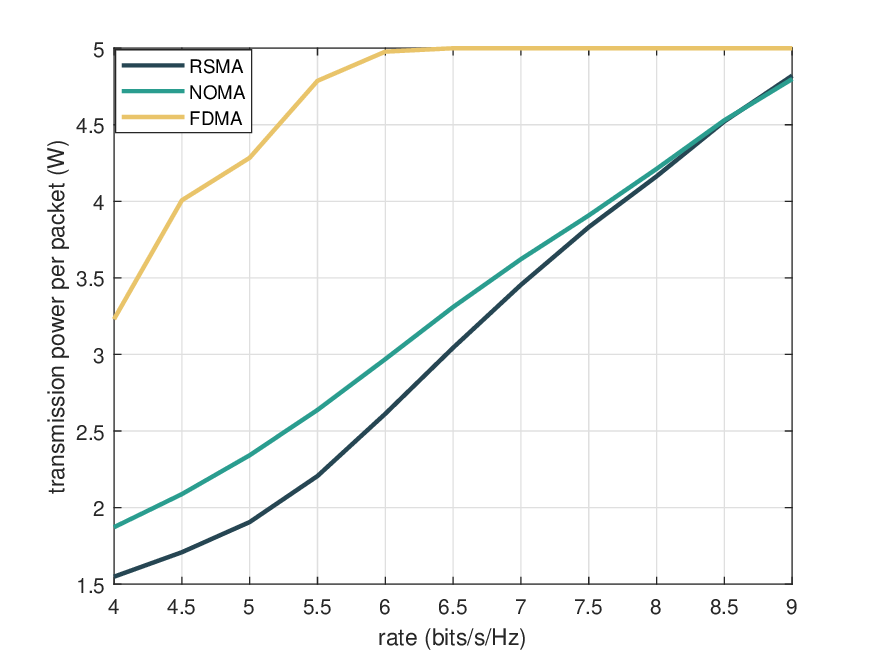}
\caption{Average power consumption of each packet in CC with 4 retransmissions.}\label{fig:T_4_CC_P}
\end{figure}

Fig. \ref{fig:T_4_CC} and Fig. \ref{fig:T_4_CC_P} present the error probabilities and average power consumption per packet with CC, respectively, and 4 retransmissions are allowed. This set can be seen as for the service which is not very sensitive to the latency. In Fig. \ref{fig:T_4_CC}, for both two users, RSMA has the best performance, NOMA has the second best, and the error probabilities in FDMA are the highest. FDMA curves are not smooth because the objective is minimizing the error probabilities, so when the rate is higher than $4.5\ \mathrm{(bit/s/Hz)}$ it allocates more resources to user 2 to minimize the error probabilities. FDMA with CC is not quite efficient when the rate is high due to the limited bandwidth while allocating resources in a non-orthogonal manner can exploit the resources better. Although NOMA has a much better performance than FDMA, for user 2 there is still room to improve. A simple example can give intuition. We assume that $s_1$ is decoded before $s_2$. Decoding $s_1$ at a relatively high rate is not always possible due to the interference from $s_2$, and decoding $s_2$ actually depends on whether decoding $s_1$ successfully, so the situation that both users cannot be decoded may happen. In RSMA, $s_{1,1}$ can be decoded and cancelled at a relatively low rate, and this increases the probability of decoding $s_2$ successfully.
Besides, the two users will stop generating new packets if either $s_{1,1}$ or $s_2$ fails. Thus, if only one stream needs to be retransmitted, it will be retransmitted without interference. When 4 retransmissions are allowed, these non-interfering copies can contribute more to achieve a higher rate. The average power consumption is shown in Fig. \ref{fig:T_4_CC_P}. FDMA consumes the most power per packet among the three schemes. In FDMA, each user occupies a limited bandwidth, so the frequency of retransmission increases. RSMA and NOMA consume less power since they do not retransmit failed streams all the time. RSMA consumes the least power when the rate is lower than $8\ \mathrm{(bit/s/Hz)}$. When the rate is higher than $8\ \mathrm{(bit/s/Hz)}$, the power consumption of RSMA and NOMA are almost the same, because RSMA tends to allocate more power to $s_{1,1}$ so that $s_2$ can experience less interference. Besides, the frequency of retransmitting $s_{1,1}$ and $s_2$ together increases. Although RSMA may boil down to NOMA, it still has an advantage in power consumption, because at $5\ \mathrm{(bit/s/Hz)}$ the error probabilities of NOMA and RSMA are around $10^{-2}$, and services may not desire an error probability which is higher than $10^{-2}$.

\begin{figure}[]
\centering
\includegraphics[scale=0.5]{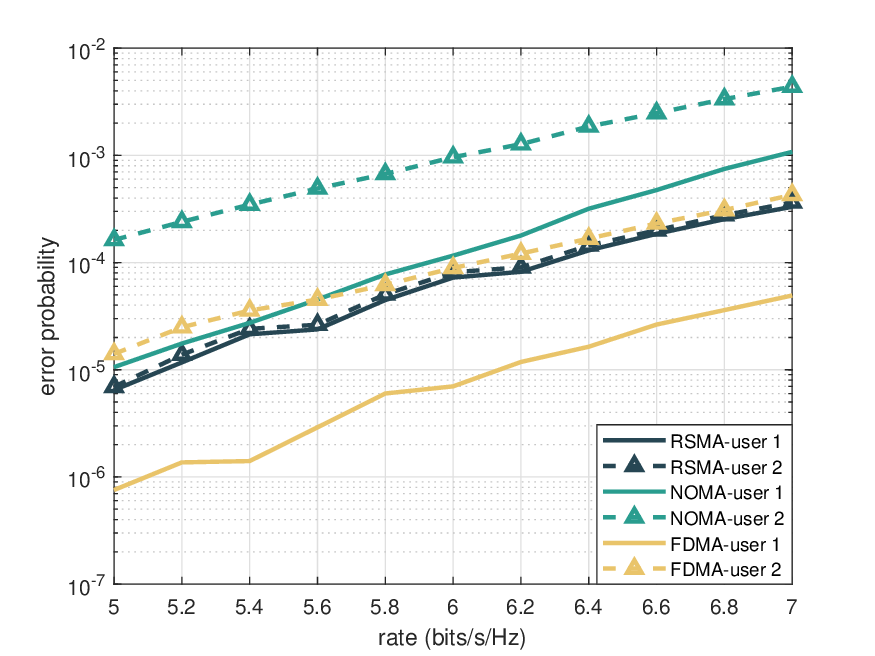}
\caption{Error probabilities of two users in IR. The average channel gains for user 1 and user 2 are $20$ dB and $15$ dB, respectively. Retransmission times $L=4$.}\label{fig:T_4_IR}
\end{figure}

\begin{figure}[]
\centering
\includegraphics[scale=0.5]{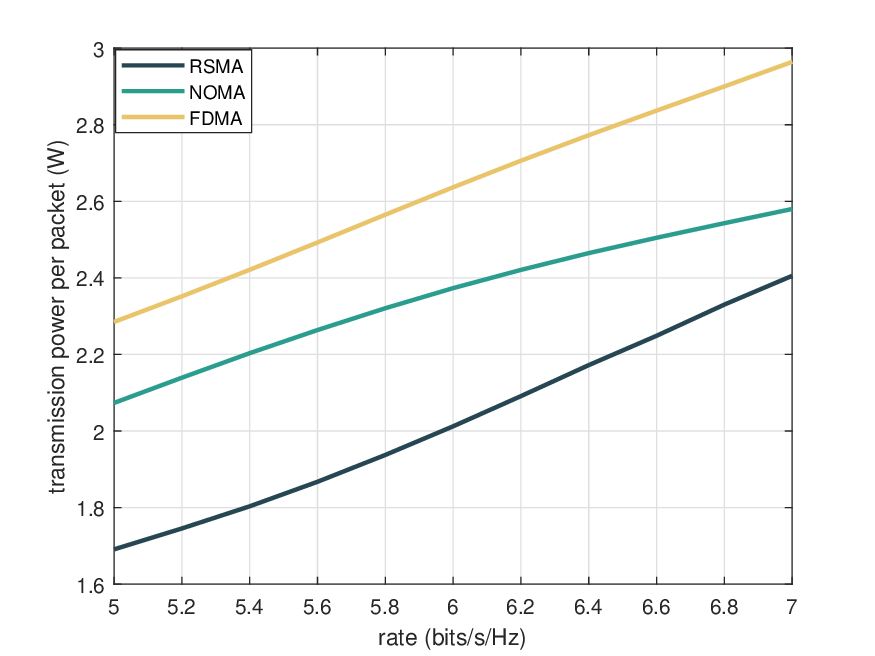}
\caption{Average power consumption of each packet in IR with 4 retransmissions.}\label{fig:T_4_IR_P}
\end{figure}

The performances when IR is applied with 4 retransmissions are shown in Fig. \ref{fig:T_4_IR} and Fig. \ref{fig:T_4_IR_P}. Compared to HARQ-CC, the achievable rate of HARQ-IR is much higher when 4 retransmissions are allowed since the logarithmic function is a concave function. The error probabilities of the two users are shown in Fig. \ref{fig:T_4_IR}. According to Fig. \ref{fig:T_4_IR}, NOMA has the highest error probabilities for both two users. While for user 1 FDMA has a lower error probability than RSMA, RSMA has a lower error probability for user 2 than FDMA. When a high rate is required and the latency requirement is not strict, it would be better to let isolated resources be dedicated to the users, because the user with good channel conditions cannot tolerate that much interference. The copies with interference would also not help much, which is also shown in Fig. \ref{fig:T_2_IR}. Hence, FDMA has lower error probabilities than NOMA though the users can only occupy part of the resources. Although users can interfere with each other in RSMA, in the retransmission rounds they do not always interfere with each other. Similar to RSMA with CC, that either $s_{1,1}$ or $s_2$ fails will cause both users to pause generating new packets, so $s_{1,1}$ or $s_2$ will not be retransmitted with the interference from a new message. In this way, the two users can share the resources but not always interfere with each other. Fig. \ref{fig:T_4_IR_P} shows the average transmission power for each packet. FDMA consumes the most power, NOMA follows, and RSMA is the lowest. For NOMA and RSMA, it is not necessary to transmit all the failed streams, so they consume less power than FDMA. RSMA uses the least power because it can mitigate the effect of interference to some extent, so fewer retransmissions are needed compared to NOMA.

According to the simulation results, RSMA with HARQ has the potential to improve the achievable rate for users who require the same service. Two users can have relatively low error probabilities with both CC and IR and different retransmission times. The simulation results also show that RSMA with HARQ has the lowest average power consumption per packet, which means it is suitable for the service of low-power devices.

\section{Conclusion and Future works}\label{sec:5}
In this work, a retransmission scheme for uplink RSMA is proposed. This scheme does not need to retransmit all the failed streams, so it avoids bringing additional complexity. We focus on the scenario that two users requiring the same service share a common BS. The decoding order at BS is set to $s_{1,1}$, $s_2$, and $s_{1,2}$, which is the optimal order to exploit the flexibility of RSMA. It is always possible for $s_{1,2}$ to be decoded with a low rate, so only $s_{1,1}$ and $s_2$ are involved with retransmitting. Power allocation between $s_{1,1}$ and $s_{1,2}$ is important and it decides which stream(s) will be retransmitted. Error probabilities for different power allocation strategies are given, and the $\alpha$ which gives the lowest sum of the error probabilities is chosen.

The simulation results present the error probabilities and the average power consumption per packet of RSMA with CC and IR, and they are compared with NOMA and FDMA. The results show that RSMA can let both users have a low error probability while consuming less energy with different retransmission times. This shows that RSMA with HARQ could be applied to services that need low latency and low power consumption. RSMA with retransmission can enhance diversity and improve efficiency, so it could be a promising scheme for future communication networks. 

Although this work focuses on the two-user case, it also provides insights for a multiple-user case. The decoding order and resource allocation need to be considered carefully, and this can be left as future work. Besides, this work aims to study the fundamental limits of the systems and assumes perfect SIC, while how imperfect SIC will impact the performance is a critical issue and it can be an interesting topic.



%

\section*{Acknowledgments}
The work of Petar Popovski was supported by the Villum Investigator Grant "WATER” from the Velux Foundation, Denmark.

\appendices
\section{}\label{FirstAppendix}
 Here the derivation of $p_{1,1}$ and $p_2$ when both $s_{1,1}$ and $s_2$ are retransmitted will be given. We let $x$ and $y$ denote the channel gains, $|h_1|^2$ and $|h_2|^2$, respectively, and their pdf follow the exponential distribution in (\ref{channel_gain_pdf}). First, let we compute (\ref{p11_origin}), and it can be rewritten as
 \begin{equation}
 \begin{aligned}
     p_{1,1}&=\int_0^{\infty} \left(\int_{\frac{\alpha x}{\gamma_{1,1}^{(1)}}-1}^{\gamma_{2}^{(1)} \left(1+\alpha x\right)} \frac{1}{\Gamma_2}e^{-\frac{y}{\Gamma_2}} dy\right) \frac{1}{\Gamma_1}e^{-\frac{x}{\Gamma_1}}dx\\&+\int_0^{\frac{\gamma_{1,1}^{(2)}}{\alpha}}\left(\int_{\gamma_{2}^{(1)} (1+\alpha x)}^{\infty}\frac{1}{\Gamma_2}e^{-\frac{y}{\Gamma_2}}dy\right)\frac{1}{\Gamma_2}e^{-\frac{x}{\Gamma_1}}dx\\
     &+\int_0^{\gamma_{2}^{(2)}}\left(\int_{\frac{\gamma_{1,1}^{(1)}(1+y)}{\alpha}}^{\infty}\frac{1}{\Gamma_1}e^{-\frac{x}{\Gamma_1}}dx\right)\frac{1}{\Gamma_2}e^{-\frac{y}{\Gamma_2}}dy.
 \end{aligned}
 \end{equation}
 The second term and third term can be computed directly, but if $\gamma_{1,1}^{(2)}$ or $\gamma_{2}^{(2)}$ is negative, which means the constraint is not applicable for the variable, it should be seen as $\infty$, since the exponential distribution is only applicable for the non-negative region. Thus, for the first term, the lower limit of the inner integral should be non-negative, so the first term can be rewritten as 
 
  \begin{equation}\label{A_1}
  \begin{aligned}
  & \int_{\frac{\gamma_{1,1}^{(1)}}{\alpha}}^{\infty}\left(\int_{\frac{\alpha x}{\gamma_{1,1}^{(1)}}-1}^{\gamma_{2}^{(1)}(1+\alpha x)}\frac{1}{\Gamma_2}e^{-\frac{y}{\Gamma_2}} dy   \right) \frac{1}{\Gamma_1}e^{-\frac{x}{\Gamma_1}} dx \\
  &+ \int_0^{\frac{\gamma_{1,1}^{(1)}}{\alpha}} \left(\int_0^{\gamma_{2}^{(1)}(1+\alpha x)}\frac{1}{\Gamma_2}e^{-\frac{y}{\Gamma_2}} dy   \right) \frac{1}{\Gamma_1}e^{-\frac{x}{\Gamma_1}} dx.
  \end{aligned}
  \end{equation}
 For the first term of (\ref{A_1}), the higher limit of the inner integral should be larger than the lower one, which can be rearranged as 
 \begin{equation}
     \left( \gamma_{2}^{(1)}-\frac{1}{\gamma_{1,1}^{(1)}}\right)\alpha  x \geq -\gamma_{2}^{(1)}-1.
 \end{equation}
 Thus, if $\gamma_{1,1}^{(1)} \gamma_{2}^{(1)} \geq1$, the integral does not change, but if $\gamma_{1,1}^{(1)} \gamma_{2}^{(1)} < 1$, $x<-\frac{\left(\gamma_{2}^{(1)}+1 \right)\gamma_{1,1}^{(1)}}{\alpha \left(\gamma_{2}^{(1)}\gamma_{1,1}^{(1)}-1 \right)}=c$, and this is the new higher limit of the outer integral of the first term of (\ref{A_1}). $p_2$ equals that $p_{1,1}$ subtracts the second term, and can be computed by the same way. Then, (\ref{p_11_s11_s2}) and (\ref{p_2_s11_s2}) can be obtained.

  \section{}
  \label{SecondAppendix}
  The derivations of (\ref{p_1}) and (\ref{p_2}) are presented. Similarly, we let $x$ and $y$ denote $|h_1|^2$ and $|h_2|^2$, respectively. The first term of (\ref{special_case_p_1}) can be represented as
  \begin{equation}
  \begin{aligned}
     \int_0^{\infty}\left(\int_{\frac{x}{\gamma_{1}}-1}^{\gamma_2(1+x)}  \frac{1}{\Gamma_2}e^{-\frac{y}{\Gamma_2}}dy\right)\frac{1}{\Gamma_1}e^{-\frac{x}{\Gamma_1}}dx.
      \end{aligned}
  \end{equation}
Here the lower limit of the inner integral should not be negative, and if it is negative, the lower limit should be $0$. Thus, it can be rewritten as
\begin{equation}\label{B1_part1}
\begin{aligned}
&\int_{\gamma_1}^{\infty}\left(\int_{\frac{x}{\gamma_{1}}-1}^{\gamma_2(1+x)}  \frac{1}{\Gamma_2}e^{-\frac{y}{\Gamma_2}}dy\right)\frac{1}{\Gamma_1}e^{-\frac{x}{\Gamma_1}}dx \\ 
&+ \int_0^{\gamma_1}\left(\int_0^{\gamma_2(1+x)}  \frac{1}{\Gamma_2}e^{-\frac{y}{\Gamma_2}}dy\right)\frac{1}{\Gamma_1}e^{-\frac{x}{\Gamma_1}}dx.
\end{aligned}
\end{equation}
Then, the second term of  (\ref{special_case_p_1}) can be represented as
\begin{equation}\label{B1_part2}
\begin{aligned}
    \int_0^{\gamma_1}\left(\int_{\gamma_2(1+x)}^{\infty} \frac{1}{\Gamma_2}e^{-\frac{y}{\Gamma_2}}dy\right)\frac{1}{\Gamma_1}e^{-\frac{x}{\Gamma_1}}dx.
\end{aligned}
\end{equation}
The sum of second term of (\ref{B1_part1}) and (\ref{B1_part2}) is
\begin{equation}
    \int_0^{\gamma_1}\frac{1}{\Gamma_1}e^{-\frac{x}{\Gamma_1}}dx=1-e^{-\frac{\gamma_{1}}{\Gamma_1}}.
\end{equation}
For the first term of  (\ref{B1_part1}), in the inner integral, the higher limit should not be less than the lower limit, so
\begin{equation}
    \left(\gamma_2-\frac{1}{\gamma_{1}}\right)x\geq-\gamma_2-1.
\end{equation}
Obviously, if $\left(\gamma_2-\frac{1}{\gamma_{1}}\right)\geq0$, there does not exist additional requirement for $x$; while if $\left(\gamma_2-\frac{1}{\gamma_{1}}\right)<0$, $x<-\frac{\left(1+\gamma_2\right)\gamma_1}{\gamma_1\gamma_2-1}$, and the first term becomes
\begin{equation}
    \int_{\gamma_1}^{-\frac{\left(1+\gamma_2\right)\gamma_1}{\gamma_1\gamma_2-1}}\left(\int_{\frac{x}{\gamma_{1}}-1}^{\gamma_2(1+x)}  \frac{1}{\Gamma_2}e^{-\frac{y}{\Gamma_2}}dy\right)\frac{1}{\Gamma_1}e^{-\frac{x}{\Gamma_1}}dx.
\end{equation}

(\ref{special_case_p_2}) can be represented as
\begin{equation}\label{partB_3}
    \begin{aligned}
    p_2&=\int_{\gamma_1}^{\infty}\left(\int_{\frac{x}{\gamma_{1}}-1}^{\gamma_2(1+x)}  \frac{1}{\Gamma_2}e^{-\frac{y}{\Gamma_2}}dy\right)\frac{1}{\Gamma_1}e^{-\frac{x}{\Gamma_1}}dx \\ 
&+ \int_0^{\gamma_1}\left(\int_0^{\gamma_2(1+x)}  \frac{1}{\Gamma_2}e^{-\frac{y}{\Gamma_2}}dy\right)\frac{1}{\Gamma_1}e^{-\frac{x}{\Gamma_1}}dx\\
    &+ \int_0^{\gamma_2}\left(\int_{\gamma_1(1+y)}^{\infty}\frac{1}{\Gamma_1}e^{-\frac{x}{\Gamma_1}}dx\right)\frac{1}{\Gamma_2}e^{-\frac{y}{\Gamma_2}}dy,
    \end{aligned}
\end{equation}
and by the same method, (\ref{p_2}) can be obtained.

\bibliographystyle{IEEEtran} 
\bibliography{references.bib}

\begin{IEEEbiography}[{\includegraphics[width=1in,height=1.25in,clip,keepaspectratio]{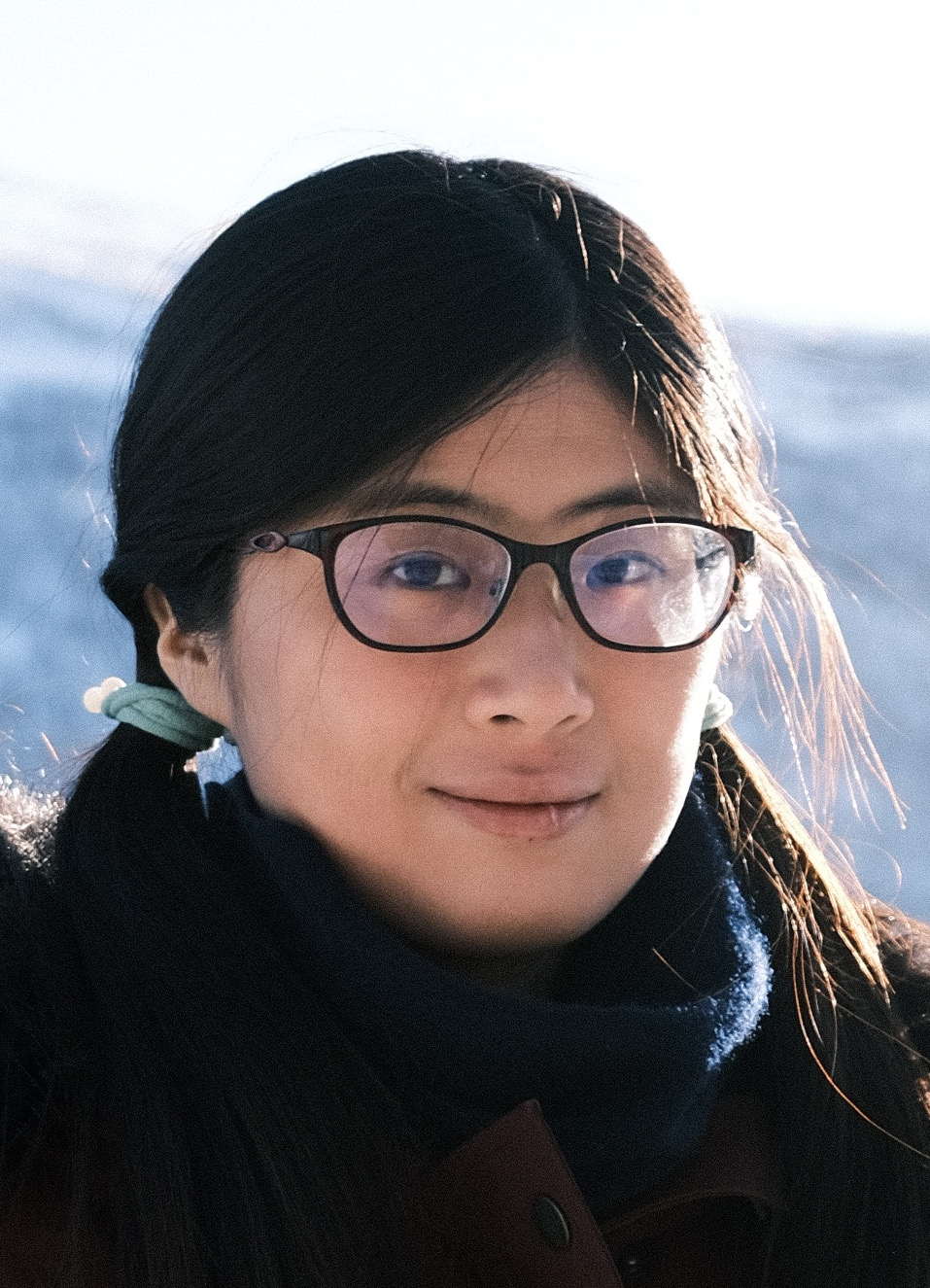}}]{Yuanwen Liu} received the BEng degree in telecommunications engineering and M.S. in electrical engineering from Beijing University of Posts and Telecommunications Beijing, China in 2019 and Northwestern University, Evanston, United States in 2021. Currently she is a PhD student with the Department of Electrical and Electronic Engineering at Imperial College London, United Kingdom. Her research interests is rate splitting multiple access. 
\end{IEEEbiography}
\vfill
\begin{IEEEbiography}[{\includegraphics[width=1in,height=1.25in,clip,keepaspectratio]{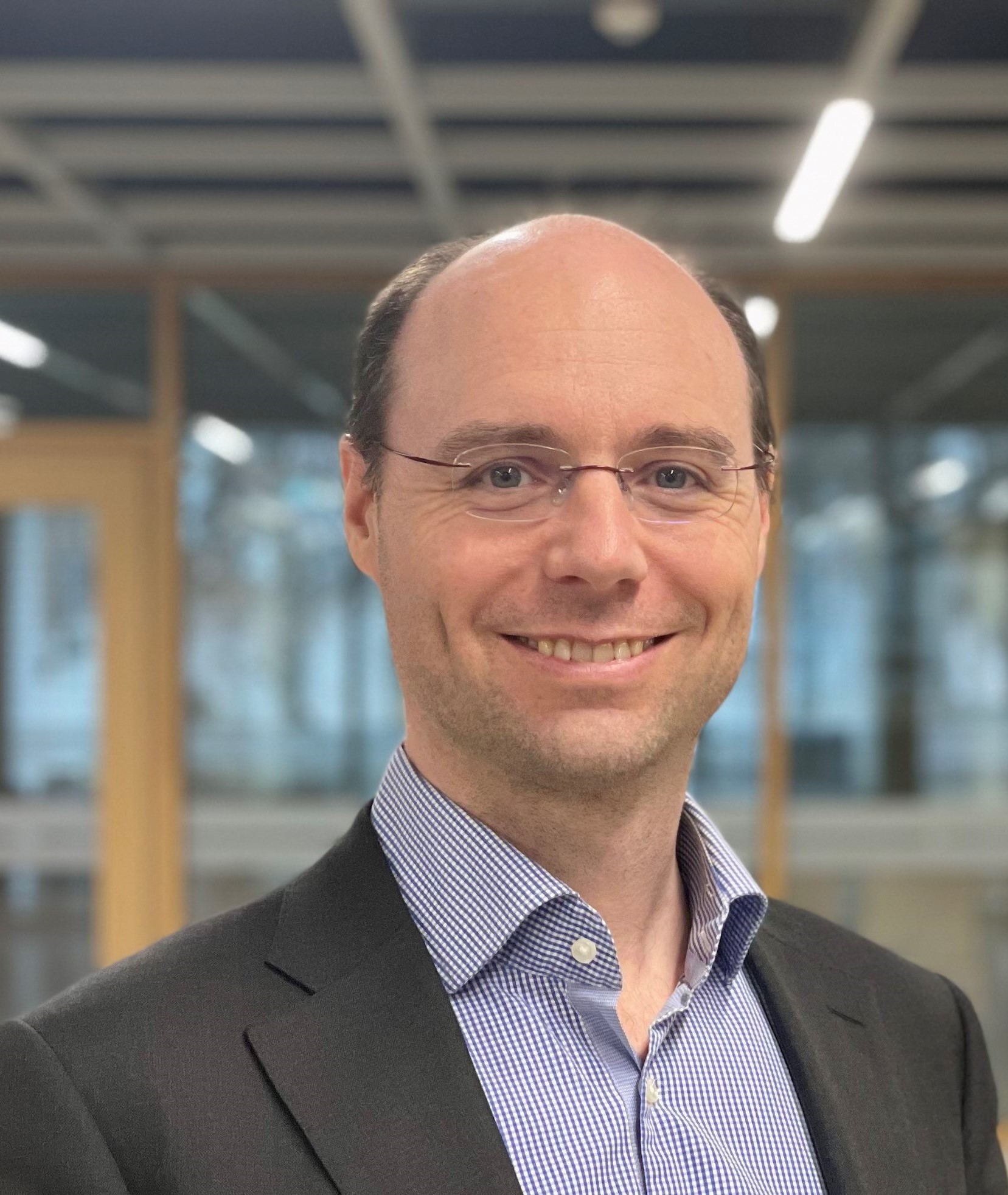}}]{Bruno Clerckx}
(Fellow, IEEE) is a (Full) Professor, the Head of the Communications and Signal Processing Group, and the Head of the Wireless Communications and Signal Processing Lab, within the Electrical and Electronic Engineering Department, Imperial College London, London, U.K. He received the MSc and Ph.D. degrees in Electrical Engineering from Université Catholique de Louvain, Belgium, and the Doctor of Science (DSc) degree from Imperial College London, U.K. He spent many years in industry with Silicon Austria Labs (SAL), Austria, where he was the Chief Technology Officer (CTO) responsible for all research areas of Austria's top research center for electronic based systems and with Samsung Electronics, South Korea, where he actively contributed to 4G (3GPP LTE/LTE-A and IEEE 802.16m). He has authored two books on “MIMO Wireless Communications” and “MIMO Wireless Networks”, 300 peer-reviewed international research papers, and 150 standards contributions, and is the inventor of 80 issued or pending patents among which several have been adopted in the specifications of 4G standards and are used by billions of devices worldwide. His research spans the general area of wireless communications and signal processing for wireless networks. He received the prestigious Blondel Medal 2021 from France for exceptional work contributing to the progress of Science and Electrical and Electronic Industries, the 2021 Adolphe Wetrems Prize in mathematical and physical sciences from Royal Academy of Belgium, multiple awards from Samsung, IEEE best student paper award, and the EURASIP (European Association for Signal Processing) best paper award 2022. He is a Fellow of the IEEE and the IET.
\end{IEEEbiography}

\begin{IEEEbiography}[{\includegraphics[width=1in,height=1.25in,clip,keepaspectratio]{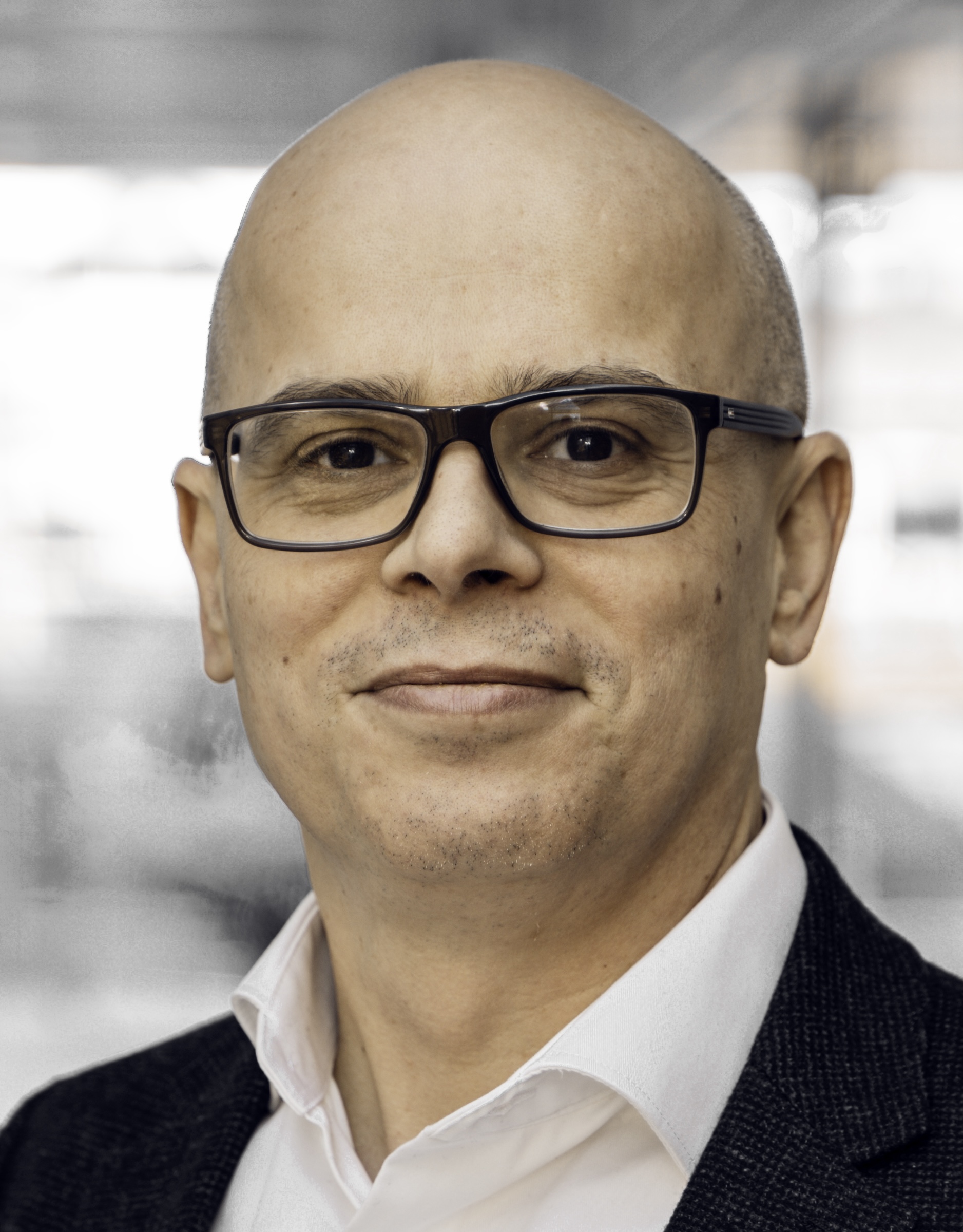}}]{Petar Popovski}
(Fellow, IEEE) is a Professor at Aalborg University, where he heads the section on Connectivity and a Visiting Excellence Chair at the University of Bremen. He received his Dipl.-Ing and M. Sc. degrees in communication engineering from the University of Sts. Cyril and Methodius in Skopje and the Ph.D. degree from Aalborg University in 2005. He received an ERC Consolidator Grant (2015), the Danish Elite Researcher award (2016), IEEE Fred W. Ellersick prize (2016), IEEE Stephen O. Rice prize (2018), Technical Achievement Award from the IEEE Technical Committee on Smart Grid Communications (2019), the Danish Telecommunication Prize (2020) and Villum Investigator Grant (2021). He was a Member at Large at the Board of Governors in IEEE Communication Society 2019-2021. He is currently an Editor-in-Chief of IEEE JOURNAL ON SELECTED AREAS IN COMMUNICATIONS and a Chair of the IEEE Communication Theory Technical Committee. Prof. Popovski was the General Chair for IEEE SmartGridComm 2018 and IEEE Communication Theory Workshop 2019. His research interests are in the area of wireless communication and communication theory. He authored the book ``Wireless Connectivity: An Intuitive and Fundamental Guide''.
\end{IEEEbiography}


\vfill


\end{document}